\documentclass[final]{ias2}

\usepackage{graphicx} 
\usepackage{multirow}
\usepackage{array} 
\usepackage{epsfig} 
\usepackage{hyperref} 

\newcommand{\lsim}{\raisebox{-0.13cm}{~\shortstack{$<$ \\[-0.07cm] $\sim$}}~} 
\newcommand{\gsim}{\raisebox{-0.13cm}{~\shortstack{$>$ \\[-0.07cm] $\sim$}}~} 
\newcommand{\beq}{\begin{eqnarray}} 
\newcommand{\eeq}{\end{eqnarray}} 
\newcommand{\tb}{\tan \beta}

\begin{document}

\markboth{Higgs Physics: Theory}{Abdelhak Djouadi}


\title{Higgs Physics: Theory\footnote{\it Talk given at  the XXV International Symposium on Lepton Photon
Interactions at High Energies \\ \hspace*{5mm}  (Lepton Photon 11), 22--27
August 2011, Mumbai, India.}}

\author[sin]{Abdelhak Djouadi} 
\email{abdelhak.djouadi@th.u-psud.fr}
\address[sin]{Laboratoire de Physique Th\'eorique, U. Paris-Sud et CNRS, 91405 Orsay Cedex, 
France}

\begin{abstract} I review the theoretical aspects of the physics of Higgs
bosons, focusing on the elements that are relevant for the production and
detection  at present hadron colliders. After briefly  summarizing the basics of
electroweak symmetry breaking in the Standard Model, I discuss Higgs production
at the LHC and at the Tevatron,  with some focus on the main production
mechanism, the gluon-gluon fusion process, and summarize the main Higgs decay
modes and the experimental detection channels. I then briefly survey the case of
the minimal supersymmetric extension of the Standard Model. In a last section, I
review the prospects for determining the fundamental properties of the Higgs
particles once they have been experimentally observed.  \end{abstract}



\maketitle

\vspace*{-10mm} 
\section{Introduction}

Establishing the precise mechanism of the spontaneous breaking of the 
electroweak gauge symmetry is  a central focus of the activity in  high energy
physics and, certainly, one of the primary undertakings of the Large  Hadron
Collider, the LHC, as well as the Tevatron. In the Standard Model (SM),
electroweak symmetry  breaking (EWSB) is achieved via the Brout--Englert--Higgs
mechanism \cite{Higgs}, wherein the neutral  component of an isodoublet scalar
field acquires a non--zero vacuum  expectation value. This gives rise to nonzero
masses for the fermions and  the electroweak gauge bosons,  which are otherwise
not allowed by the  ${\rm SU(2)\!\times\! U(1)}$ symmetry. In the sector of the
theory with broken symmetry, one of the four degrees of freedom of the original
isodoublet field, corresponds to a physical particle:  the scalar Higgs  boson
with  ${\rm J^{PC}}\!=\!0^{++}$ quantum numbers under parity and charge
conjugation.  The Higgs couplings to the fermions and gauge bosons are  related
to the  masses of these particles and are thus decided by the symmetry breaking
mechanism.  For detailed reviews of the Higgs properties, see
Refs.~\cite{HHG,Djouadi:2005gi}

In contrast, the mass of the Higgs boson itself, although expected  to be in the
vicinity of the EWSB scale $v\!\approx\! 250$ GeV,  is undetermined.  Before the
start of the LHC, one available direct information on this parameter was the
lower limit  $M_H \gsim 114.4$ GeV at 95\% confidence level (CL) established at
LEP2 \cite{Barate:2003sz}.  Very recently,  the Tevatron has collected a large
data set which allowed the CDF and D0  collaborations to be sensitive to a
SM--like Higgs particle and,  indeed, the mass range between 156 GeV and 177 GeV
has been excluded,  again at the 95\% CL \cite{Tev-exclusion}. Furthermore, the
high accuracy of the electroweak data measured at LEP, SLC and the Tevatron
\cite{PDG}  provides an indirect sensitivity to $M_H$:  the Higgs boson
contributes logarithmically, $\propto \log (M_H/M_W)$, to the radiative
corrections to the $W$ and $Z$ boson propagators. A global fit of the
electroweak precision data yields the value $M_H=92^{+34}_{-26}$ GeV,
corresponding to a 95\% CL upper limit of $M_H \lsim 161$ GeV \cite{LEPEWWG}.
Another analysis, using a different  fitting  program gives a comparable  value
$M_H=96 ^{+31}_{-24}$ GeV \cite{Gfitter}. In both cases,  the Higgs mass values
given above are when the limits from direct searches are not included in the
global fits.

From the theoretical side, the presence of this new weakly coupled degree of
freedom is a crucial ingredient  for a unitary electroweak theory. Indeed, the
SM without the Higgs particle is not self-consistent at high energies as it
leads  to scattering amplitudes of the massive electroweak gauge  bosons that
grow with the square of the center of mass energy and perturbative unitarity
would be lost at energies above the  TeV scale.  In fact, even in the presence
of a Higgs boson, the $W$ and $Z$ bosons could  interact very strongly with each
other and, imposing the unitarity requirement in the $W$ and $Z$ boson
high--energy scattering amplitudes leads to the important Higgs mass bound $M_H
\lsim 700$ GeV \cite{H-LQT}, implying that the particle is kinematically
accessible at the LHC. It is interesting  to note, as an aside, that just the
requirement of perturbative  unitarity in these scattering amplitudes leads to a
model with exactly  the same particle content and couplings as the
SM~\cite{cornwall}.

Clearly, the discovery of this last missing piece of the SM is a matter of
profound importance.  In fact, in spite of its phenomenal success  in explaining
the precision data~\cite{PDG},  the SM can not be considered to be established 
completely until the Higgs particle is  observed experimentally and, further,
its fundamental properties such as  its mass, spin and other quantum numbers, as
well as its couplings to  various matter and gauge particles and its
self-couplings are established. These studies are  important not only to crown
the SM as the correct theory of fundamental particles and interactions among
them, but also to  achieve further clarity into the dynamics of the EWSB
mechanism.  The many important questions which one would like answered are: does
the dynamics involve new strong interactions and is the Higgs a composite field?
if elementary Higgs particles indeed exist  in nature, how many fields are there
and in which gauge representations do  they appear? does the EWSB sector involve
sizable CP violation? etc. 

Theoretical realizations span a wide range of scenarios extending from weak  to
strong breaking mechanisms, including the so called Higgsless theories in extra
dimensional models. As far as the representations of the gauge group  are
concerned, there is again a whole range starting from models involving  light
fundamental Higgs fields, arising from an SU(2) doublet,  such as in the  SM and
its supersymmetric extensions which include two--Higgs doublets in the minimal
version, the MSSM, to those containing additional singlet fields or higher 
representations in extended versions in unified theories and/or alternative
theories such as little Higgs models.  Furthermore, the link between particle
physics and cosmology means that the EWSB mechanism can have implications for
the  generation of the baryon asymmetry in the early universe and could play an
important role in the annihilation of the new particles that are responsible for
the cosmological dark matter and thus impact their density in the universe
today.   An understanding of the EWSB mechanism at a more fundamental level
might also hold clues about why the three generations of quarks and leptons have
masses which differ from each other; the so called flavour issue.  A complete
discussion of Higgs physics thus touches upon almost all the issues under active
investigation in theoretical  and experimental particle physics. 

In this talk, I will discuss the physics of Higgs bosons focusing primarily on
theoretical aspects related to searches at the Tevatron and the LHC;  the
experimental aspects are discussed  in the talks of the Tevatron and LHC 
collaborations to the conference \cite{Tevatron,Nisati,Sharma}. I will consider 
in some details the SM Higgs case and only briefly survey the case of the Higgs
particles of supersymmetric theories. However, in this case, due to the lack of
space/time, I will only discuss the CP conserving MSSM; an account of the 
phenomenology of the CP--violating MSSM, the next--to--MSSM as well as other
supersymmetric and non--supersymmetric extensions, can be found, for instance,
in Refs.~\cite{CPV-MSSM,NMSSM,alternatives}.  More theoretical aspects of  EWSB
in  extensions of the SM have been  discussed  at this conference by G.
Bhattachryya   \cite{Gautham} and  M. Peskin  in his summary talk \cite{Peskin}.

\section{The SM Higgs boson at hadron colliders}

We summarize here the rates for the main Higgs production mechanisms at hadron
colliders, including the higher order radiative corrections and the associated 
theoretical uncertainties,  as well as the decay and  detection channels,
focusing on  the SM Higgs case.\vspace*{-3mm} 

\subsection{The SM Higgs production cross sections}

\begin{figure}[!h]
\vspace*{-.2cm}
\begin{center}
\vspace*{-1.cm}
\hspace*{-1.5cm}
\includegraphics*[width=9cm,height=5cm] {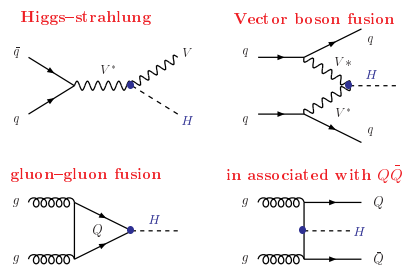}
\end{center}
\vspace*{-1.1cm}
\caption{Feynman diagrams for the leading production mechanisms of the SM 
Higgs boson at hadron colliders.}
\label{ppmecha-lhc}
\vspace*{-.2cm}
\end{figure}

There are essentially four mechanisms for the single production of the SM Higgs
boson at hadron colliders; some Feynman diagrams are shown in 
Fig.~\ref{ppmecha-lhc}.  

\begin{figure}[!h]
\vspace*{-.6cm}
\hspace*{-11mm}
\mbox{
\includegraphics[scale=0.38]{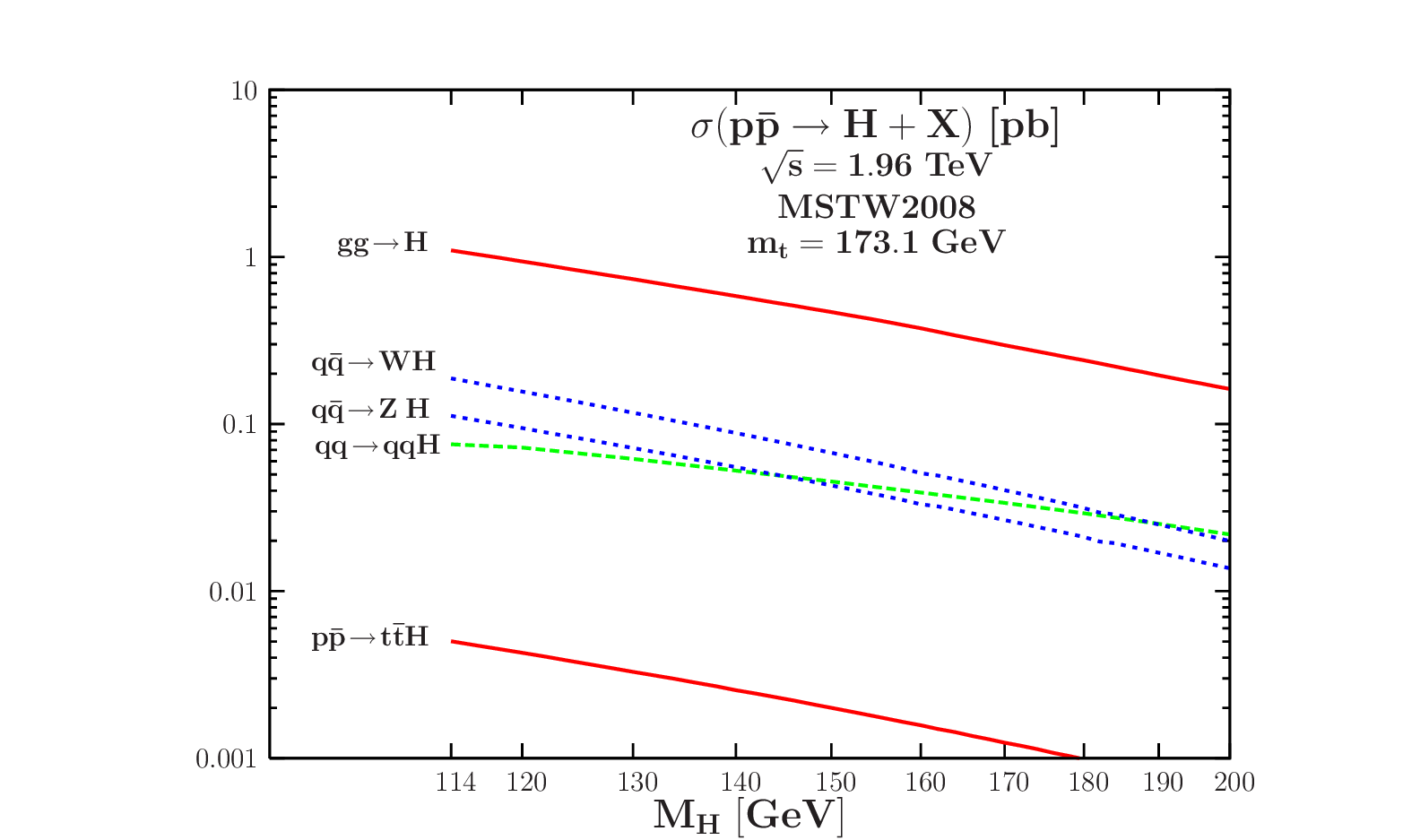}\hspace*{-5mm}
\includegraphics[scale=0.42]{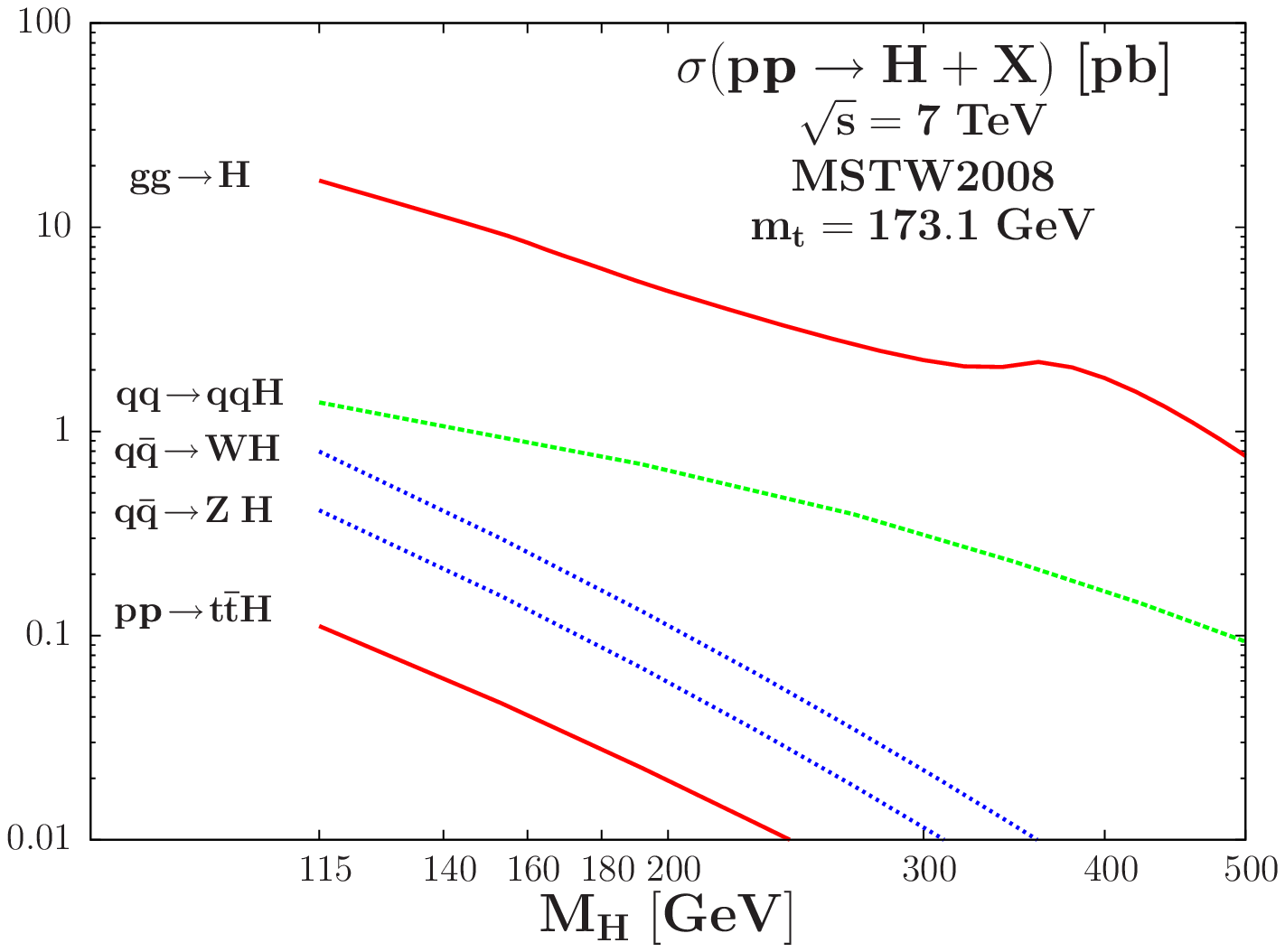}
}
\vspace*{-.5cm}
\caption{The production cross sections for the SM Higgs boson at the 
Tevatron  and early LHC  as a function of $M_H$ in the main channels. 
From Refs.~\cite{pap-tev,pap-lhc}.}
\label{ppproduct-lhc}
\vspace*{-.1cm}
\end{figure}

The total production cross sections, borrowed from Refs~\cite{pap-tev,pap-lhc} 
and obtained using adapted versions of the  programs of Ref.~\cite{Michael}, are
displayed in Fig.~\ref{ppproduct-lhc} for  the  Tevatron with $\sqrt s=1.96$ TeV
and the LHC with $\sqrt{s}=7$ TeV as a function of the Higgs mass; the top quark
mass is set to $m_t=173.1$ GeV \cite{PDG} and the MSTW \cite{PDF-MSTW} parton
distributions functions  (PDFs) have been adopted. The most important higher
order QCD and electroweak  corrections, summarized below for each
production channel, have been implemented .\smallskip

The gluon--gluon fusion process $gg \to H$ is by far the dominant
production channel for SM--like Higgs particles at hadron colliders. The process, which
proceeds through triangular heavy quark loops, has been proposed in the late
1970s in Ref.~\cite{ggH-LO} where the  $ggH$ vertex and the production cross
section have been derived. In the SM, it is dominantly mediated by the top quark
loop contribution, while the bottom quark contribution does not exceed the
10\% level at leading order. This process is known to be subject to extremely 
large QCD radiative corrections that can be described by an associated
$K$--factor defined as the ratio of the higher order (HO) to the lowest order
(LO) cross  sections, consistently evaluated with the value of the strong
coupling $\alpha_s$ and the parton distribution  functions  taken at the
considered perturbative order.   

The next-to-leading-order (NLO) corrections in QCD have been calculated in the
1990s. They are known both  for infinite \cite{ggH-NLO} and finite
\cite{ggH-NLO-ex,SDGZ} loop quark masses and, at the $\sqrt{s}\!=\!7$ TeV LHC,
lead to a $K$--factor $K_{\rm NLO}\sim 1.8$ in the low Higgs mass range, if the
central scale of the cross section is chosen to be $M_H$.  It has been shown in
Ref.~\cite{SDGZ} that working in an effective field theory (EFT) approach in
which the top quark mass is assumed to be infinite is a very  good approximation
for  $M_H \lsim 2m_t$, provided that the leading order cross section contains
the full  $m_t$ and $m_b$ dependence. The challenging calculation of the
next-to-next-to-leading-order (NNLO) contribution has been done  \cite{ggH-NNLO}
only in the EFT approach with $M_H \ll 2m_t$  and, at $\sqrt s=7$  TeV, it leads
to a $\approx 25\%$ increase of the cross section,  $K_{\rm NNLO} \sim 2.5$. The
resummation of soft gluons is known up to next-to-next-to-leading-logarithm
(NNLL) and, again, increases the cross section by slightly less than 10\%
\cite{ggH-resum,ggH-FG}. The effects of soft--gluon resumation at NNLL can be
accounted for in $\sigma^{\rm NNLO}(gg\to H)$ by lowering the central value of
the  renormalization and factorization scales, from 
$\mu_0\!=\!\mu_R\!=\!\mu_F\!=\!M_H$ to $\mu_0\!=\!\frac12 M_H$; see e.g.
Ref.~\cite{ggH-radja}. The latter value is   chosen for the cross section at
NNLO displayed in Fig.~\ref{ppproduct-lhc}.   Note in passing that the choice
$\mu_0=\frac12 M_H$  also improves the convergence of the perturbative series
and is more appropriate to describe the kinematics of the process. Some small
additional corrections beyond NNLO have been also calculated
\cite{Moch,ggH-HO,ggH-HO-mt}. The electroweak corrections  are known both in
some approximations \cite{ggH-EW} and exactly at NLO \cite{ggH-actis} and
contribute at the level of a few percent; there  are also small mixed NNLO
QCD--electroweak effects which have  been calculated  in an effective approach
valid for $M_{H}\ll M_W$ \cite{ggH-radja}.

\begin{figure}[!h]
\vspace*{-.2cm}
\begin{center}
\includegraphics*[width=8.5cm,height=5.3cm]{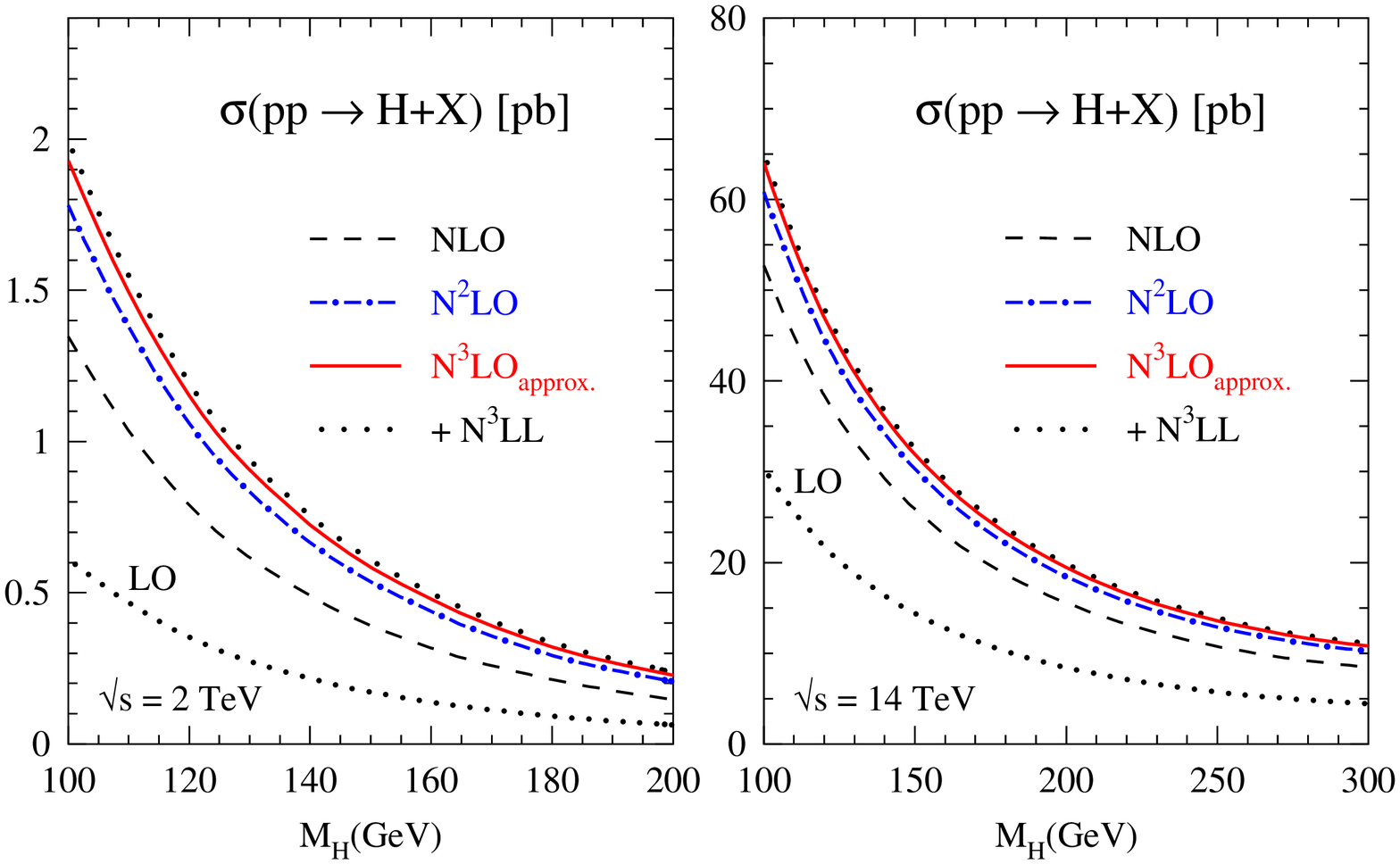} 
\end{center}
\vspace*{-.5cm}
\caption{SM Higgs production cross sections in the $gg$ fusion  process at the
14 TeV LHC and the Tevatron at the various perturbative
orders in QCD. From Ref.~\cite{Moch}.} 
\label{Kfact}
\vspace*{-.3cm}
\end{figure}

The  QCD corrections to $gg\to H$ at $\sqrt s=7$ TeV are  smaller than the
corresponding ones at the Tevatron as the $K$--factors in this case are $K_{\rm
NLO} \approx 2$ and $K_{\rm NNLO} \approx 3$ (with again a central scale equal
to $M_H$). In turn, at the LHC with $\sqrt s=14$ TeV, the $K$--factors  are 
smaller, $K_{\rm NLO} \approx 1.7$ and $K_{\rm NNLO} \approx 2$. The
perturbative series shows thus a better (converging) behavior at LHC than at
Tevatron energies.  The impact of all these QCD corrections is summarized  in
Fig.~\ref{Kfact} for the LHC at $\sqrt s=14$ TeV and for the  Tevatron.
Updates of the Higgs cross sections including the relevant higher order
corrections have been performed in various recent papers, see e.g.  
Refs.~\cite{ggH-FG, ggH-radja,pap-tev,pap-lhc,LHCXS,other-updates}.

Note that the QCD corrections to the differential  distributions, and in
particular to the Higgs transverse momentum and rapidity distributions, have
also been  calculated at NLO (with a resummation for the former) and shown to be
in general rather large; see Ref.~\cite{LHCXS-2} for an account.  These 
calculations are implemented in several programs  \cite{ggH-distr} which allow
to derive the full differential cross section and hence the  $gg\! \to\! H$
cross section with cuts. 

Let us now briefly discuss the other Higgs production processes at hadron
colliders.

The Higgs--strahlung processes  where the Higgs is produced in association with
$V=W,Z$  bosons, $q\bar q \to HV$ \cite{HV-LO}, are known exactly up to NNLO in
QCD \cite{HV-NLO,DS,HV-NNLO} (at NLO it can be inferred from Drell--Yan
production)  and up to NLO for the electroweak corrections \cite{HV-EW}. In
Fig.~2, the corrections  are evaluated at a central scale  $\mu_0= M_{HV}$, i.e.
the invariant mass of the $HV$ system. Here, the QCD $K$--factors are moderate
both at the Tevatron and the LHC, $K_{\rm NNLO} \sim 1.5$ and the electroweak 
corrections reduce the cross section by a few percent. The remaining scale
dependence is very small, making this process the theoretically cleanest of all
Higgs production processes. Note that the differential cross section is also
known up to NNLO; see for instance  Ref.~\cite{HV-dist}.

The  vector boson fusion channel, where the Higgs is produced in association
with two jets,  $qq \to Hqq$ \cite{VVH-LO}, is the second most important process
at the LHC. The QCD corrections, which can be obtained in the
structure--function approach, are moderate at NLO being at the level of 10\%
\cite{VVH-NLO,DS}. The NNLO corrections which have been calculated recently 
have  been found to be very small \cite{VVH-NNLO}. The  electroweak corrections
have been derived in Ref.~\cite{VVH-EW}. The corrections including cuts, and in
particular corrections to the transverse momentum and rapidity distributions,
have also been calculated  and implemented into a parton--level Monte--Carlo
programs such as  in Ref.~\cite{VV-MC}.

Finally, there is associated Higgs production with top quark pairs \cite{ttH-LO}
which is only relevant at the LHC.  The cross  section is rather involved at
tree--level since it is a three--body process, and the calculation of the NLO
corrections was a real challenge which was met  a few years ago \cite{ttH-NLO}.
The $K$--factors turned out to be rather small, $K\sim 1.2$ and $K \sim 1$ at
the LHC with, respectively,   $\sqrt s=14$ TeV and 7 TeV ($K\sim 0.8$ at the
Tevatron) if the central scale is chosen to be $\mu_0\!=\!\frac12(M_H+2m_t)$. 
However, the scale dependence is drastically reduced from a factor two at LO to
the level of 10--20\% at NLO.

Note that there are also other Higgs production processes at hadron colliders
but which are of higher perturbative order. For instance double Higgs
production, which is sensitive to the Higgs self--coupling, can be produced in
various processes such as $gg \to HH$,  but with rather low cross sections at
the LHC as will be discussed later.

\subsection{Theoretical uncertainties on the cross section}

It is well known  that the production cross sections at hadron colliders as well
as the associated  kinematical distributions are  generally affected by various
theoretical uncertainties. In the case of the gluon--gluon fusion channel, 
these theoretical uncertainties turn out to be particularly large. They are
stemming from three main sources. 

\begin{figure}[hbtp]
\begin{center}

\epsfig{file=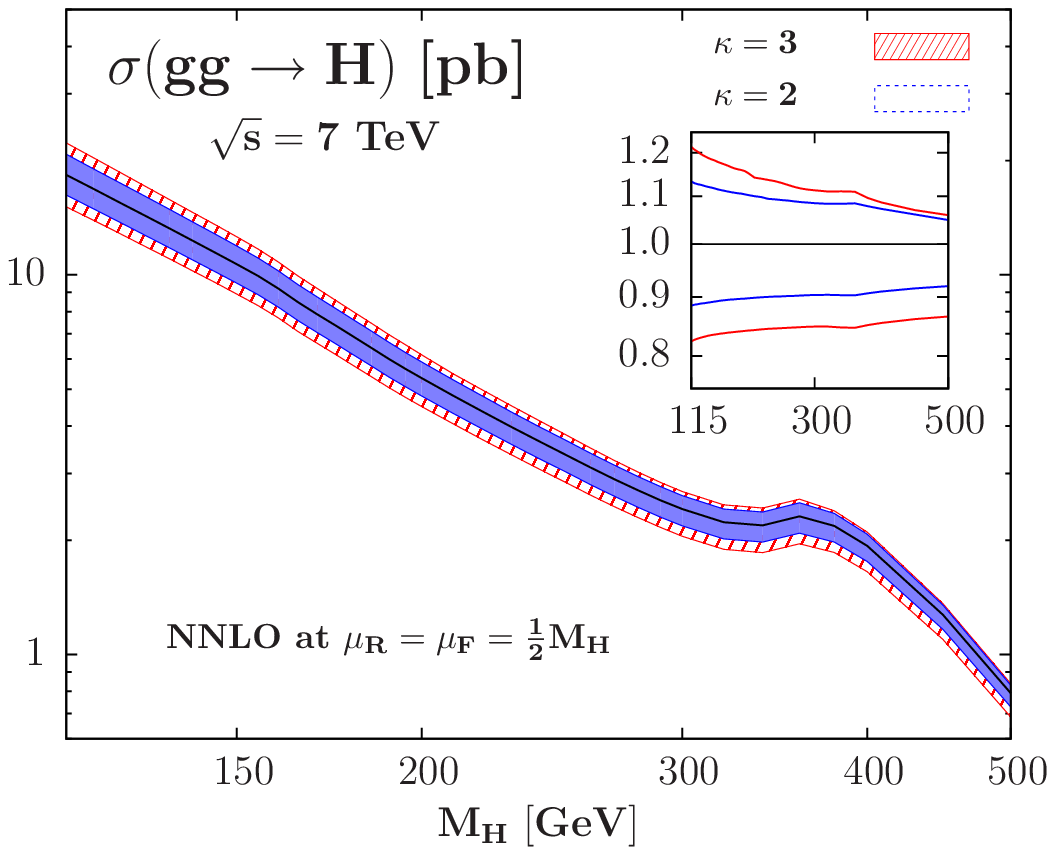,width=4cm}
\epsfig{file=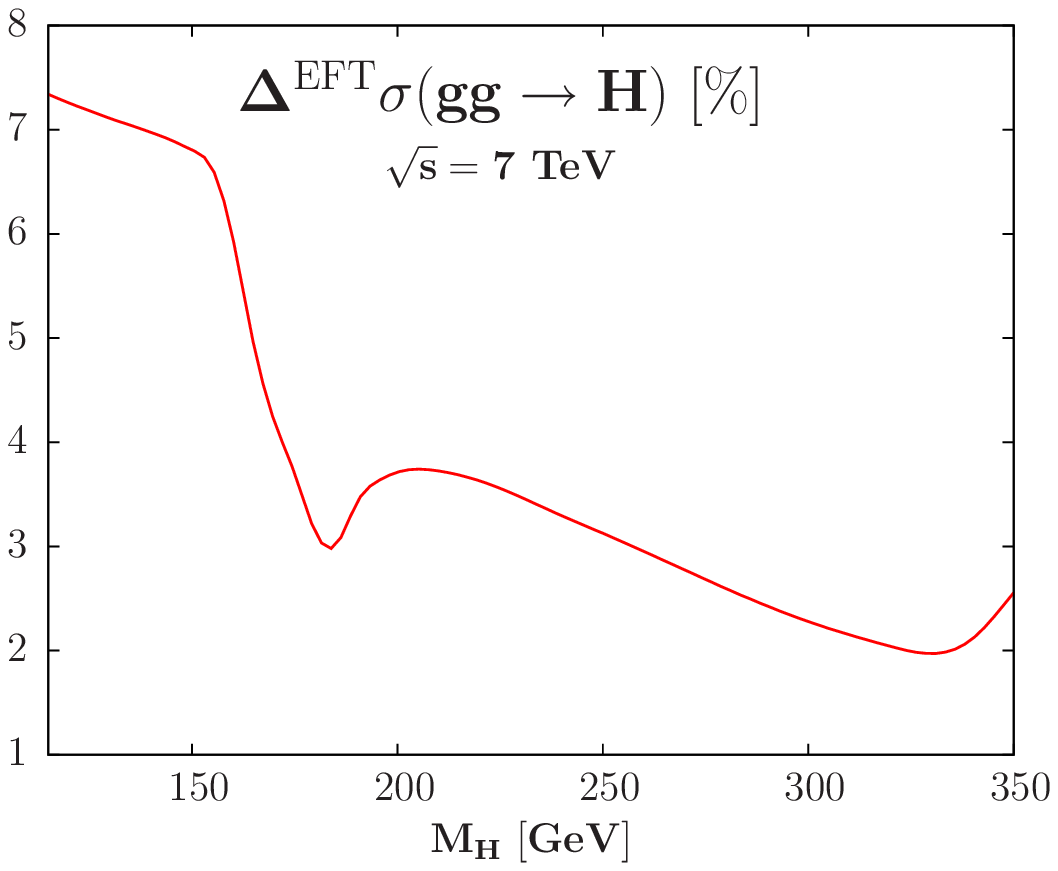, scale=0.38} 
\epsfig{file=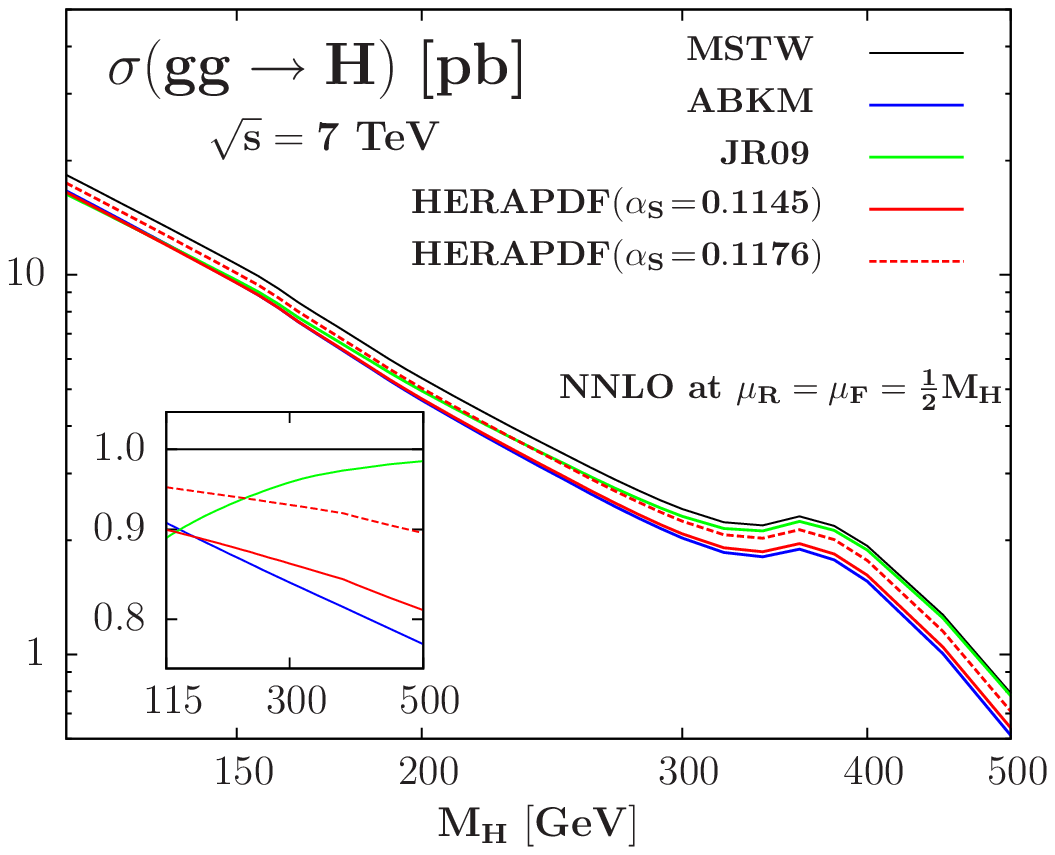,width=4cm}
\end{center}
\vspace*{-5mm}
\caption{The theoretical uncertainties on $\sigma_{gg\to H}^{\rm NNLO}$ 
at the LHC with $\sqrt s=7$ TeV as a function of $M_H$ from scale variation
(left), the use of the EFT approach (center) and when using different NNLO
PDFs; from Ref.~\cite{pap-lhc}.}
\label{errors}
\vspace*{-2mm}
\end{figure}

First, the perturbative QCD corrections to the $gg\to H$ cross section are so
large that one may question the reliability of the perturbative series (in
particular, at the Tevatron as discussed above)  and the  possibility of  still
large  higher  order contributions beyond NNLO cannot be excluded. The effects
of the unknown  contributions are usually estimated from the variation of the
cross section  with the renormalisation $\mu_R$ and  factorisation $\mu_F$
scales at which the process is evaluated.  Starting from a median scale taken to
be $\mu_R\!=\!\mu_F\!=\!\mu_0\! =\! \frac12 M_H$ in the $gg\! \to\! H$ process,
the current convention is to vary these two scales within the range
$\mu_0/\kappa \!\le \!\mu_R, \mu_F \!\le \! \kappa \mu_0$ with the choice
$\kappa\!=\!2$. This leads to a $\approx \pm 10\%$ uncertainty at the 7 TeV LHC
\cite{LHCXS,pap-lhc} as can be seen in   Fig.~\ref{errors} (left). 

Another problem that is specific to the $gg \to H$ process is that, already at
LO, it occurs at the one--loop level with the additional complication of having
to account for the finite mass of the  loop particle.  This renders the NLO
calculation extremely complicated and the NNLO calculation a formidable  task.
Luckily, one can work in an effective field theory (EFT) approach in which the
heavy loop particles are integrated out, making the calculation of the
contributions beyond NLO possible. While this approach is justified for the
dominant top quark contribution for $M_H\! \lsim\! 2m_t$ \cite{ggH-HO-mt}, it is
not valid for the $b$-quark loop and for those involving the electroweak gauge
bosons \cite{ggH-EW}. The uncertainties induced by the use of the EFT approach
at NNLO are estimated to be  of ${\cal O}(5\%)$ \cite{pap-lhc}. 

A third problem is due to the presently not satisfactory  determination of the
parton distribution  functions (PDFs). Indeed,  in this $gg$ initiated process,
the gluon  densities are poorly constrained, in particular in the high
Bjorken--$x$  regime (which is particularly relevant for the Tevatron).  
Furthermore, since $\sigma^{\rm LO}_{gg\!\to \!H} \propto \alpha_s^2$ and
receives large contributions  at ${\cal O}(\geq \alpha_s^3)$, a small change of
$\alpha_s$ leads to a large variation of $\sigma^{\rm NNLO}_{gg \to H}$.
Related  to that  is  the significant difference between the world average
$\alpha_s$ value and the one from deep-inelastic  scattering (DIS) data used in
the PDFs \cite{PDG}.  There is a statistical method to estimate the PDF 
uncertainties by allowing  a 1$\sigma$ (or more) excursion of the experimental
data that are used to perform the global fits. In addition, the MSTW
collaboration \cite{PDF-MSTW} provides a scheme that allows for a combined
evaluation of the PDF uncertainties and the (experimental and theoretical) ones
on $\alpha_s$. In Ref.~\cite{pap-lhc}, the  combined 90\% CL PDF+$\Delta^{\rm
exp}\alpha_s+\Delta^{\rm th}\alpha_s$ uncertainty on  $\sigma_{gg\to H}^{\rm
NNLO}$ at the 7 TeV LHC was found to be of order 10\%. However, this  method
does not account for the theoretical assumptions that enter into the
parametrization of the PDFs;   a way to access this theoretical uncertainty is
to compare the results for the central values of the cross section with the
best--fit PDFs when using  different parameterizations \cite{NNLO-PDFs} as shown
in the right panel of Fig.~\ref{errors}.

\begin{figure}[!h]
\vspace{-.1cm}
\begin{center}
\mbox{
\includegraphics*[width=6cm,height=5cm]{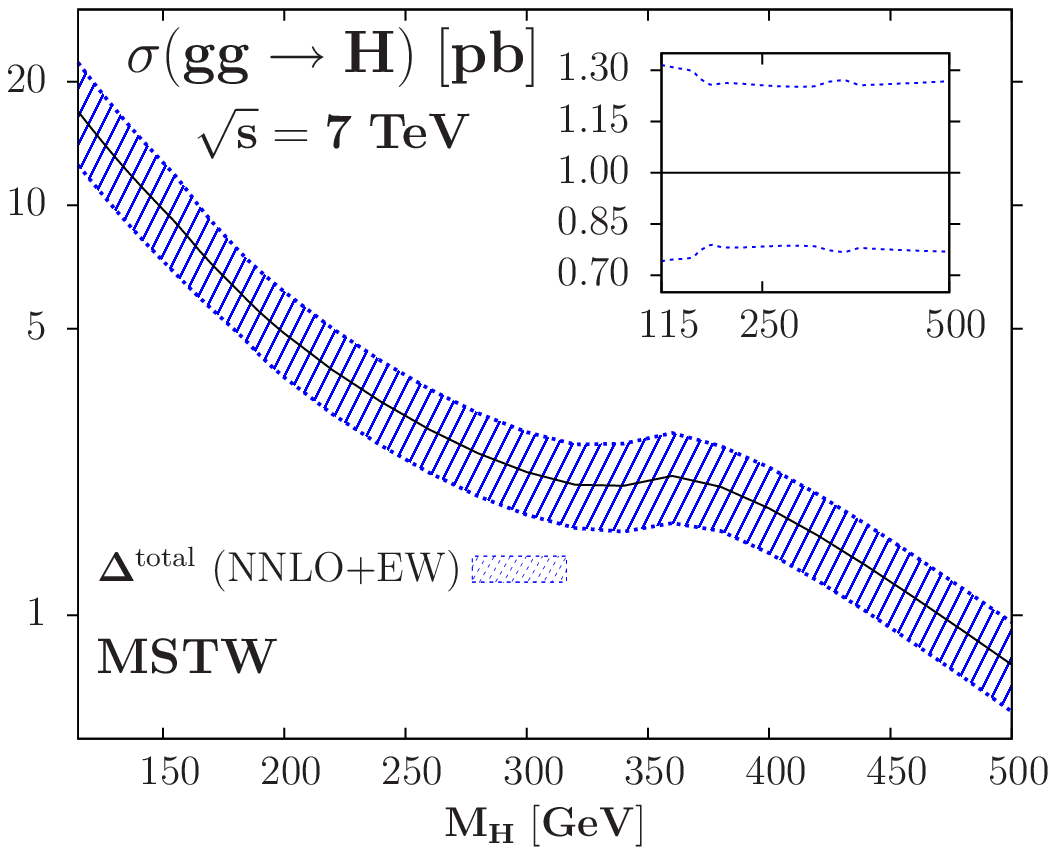}\hspace{2mm}
\includegraphics*[width=6cm,height=5cm]{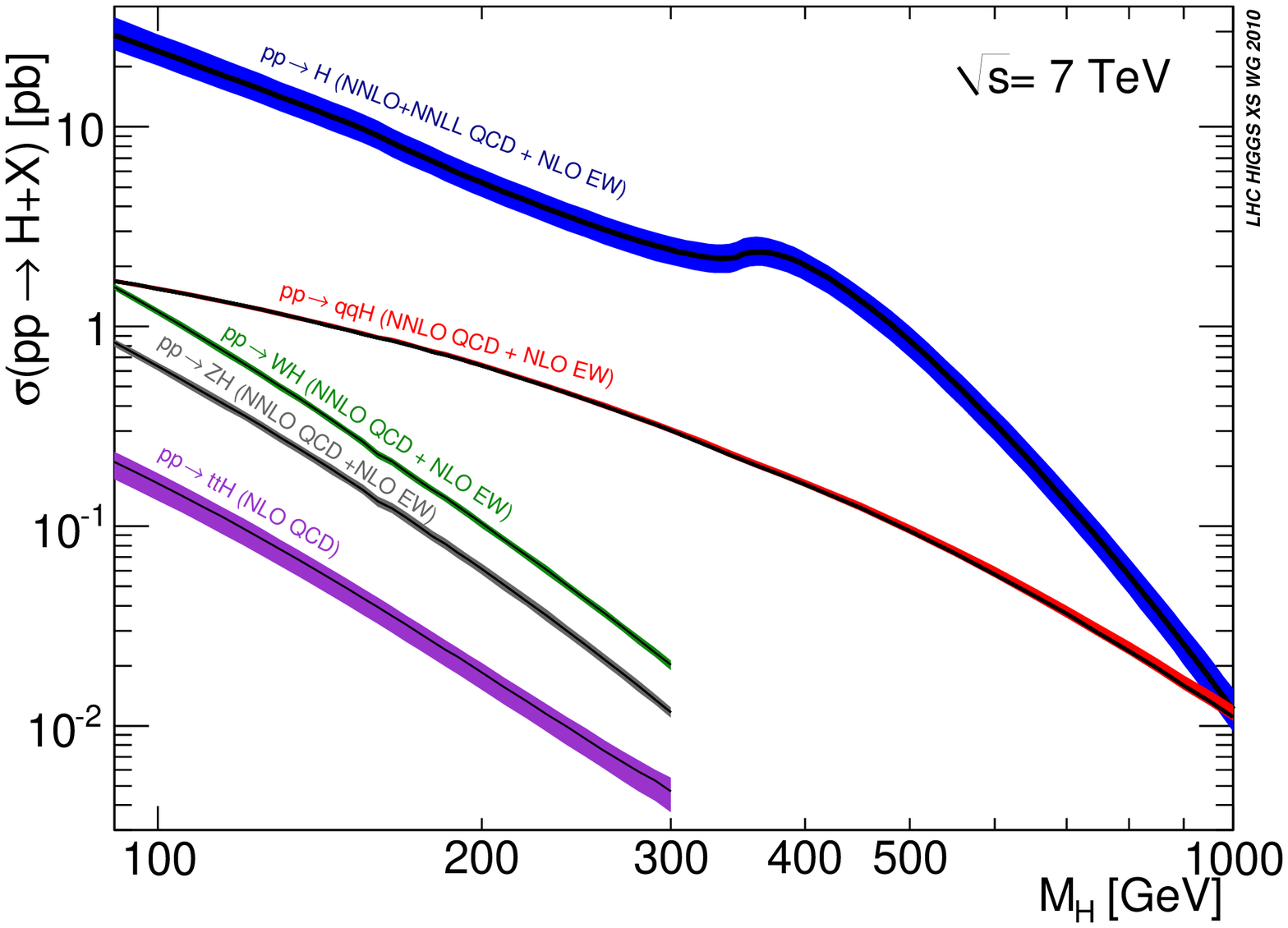} 
}
\end{center}
\vspace{-.4cm}
\caption{The total uncertainties in the Higgs production cross sections
at the 7 TeV LHC as a function of $M_H$: left is for $gg\to H$ \cite{pap-lhc} 
and right is for all processes \cite{LHCXS}.}
\label{errors-all}
\vspace*{-.3cm}
\end{figure}

An important issue  is the way these various uncertainties should be combined.
As advocated in Refs.~\cite{pap-lhc,LHCXS}, one should be conservative  and
add all these uncertainties linearly (this is equivalent of assuming that the
PDF uncertainty is a pure theoretical uncertainty with a flat prior). In the
mass range below $M_H \lsim 200$ GeV,  this would lead to an uncertainty of
about 20--25\% on $\sigma^{\rm NNLO}_{gg\to H}$ as exemplified in the left-hand
side of Fig.~\ref{errors-all}.  

In the case of the three other Higgs production channels, because the QCD
corrections are moderate, the theoretical uncertainties are much smaller: at the
LHC with $\sqrt s=7$ TeV, they are at level of $\approx 5\%$ for the
Higgs--strahlung and vector boson fusion processes and $\approx 15\%$ in the
case of $t\bar t H$ production as exemplified in the right--hand side of 
Fig.~\ref{errors-all}.

Note that at the Tevatron where the QCD corrections in the $gg \to H$ process
are  larger than at the LHC, it is wise to extend the domain of scale variation
and adopt instead a value  $\kappa\!=\!3$. This is the choice made in
Ref.~\cite{pap-tev} which resulted in a ${\cal O}( 20\%)$ scale uncertainty. In
addition, because one is in a higher Borjken--$x$ region at the Tevatron, the 
gluon density is less constrained and the PDF uncertainty is also larger than at
the LHC. The total  theoretical uncertainty on $\sigma_{gg\to H}^{\rm NNLO}$ at
the Tevatron is  estimated to be twice as large as at the LHC \cite{pap-tev}. 
In turn, for the Higgs--strahlung process, it is of ${\cal O}(5\%)$ only.

\subsection{Higgs decay modes}

Once its mass is fixed the profile of the Higgs particle is uniquely
determined and its production rates and decay widths are fixed.  As
its couplings to different particles are proportional to their masses, 
the Higgs boson  will have the tendency to decay into the heaviest particles
allowed by phase space. The Higgs decay modes and their branching
ratios are briefly summarized below.

In the ``low--mass" range, $M_H \lsim 130$ GeV,  the Higgs boson decays into  a
large variety of channels. The main mode is by far the decay into  $b\bar{b}$
with a $\sim$ 60--90\% probability followed by the decays into $c\bar{c}$ and 
$\tau^+\tau^-$ with $\sim$ 5\% branching ratios. Also of significance is the top--loop 
mediated decay into gluons, which occurs at the level of $\sim$ 5\%.  The top
and $W$--loop mediated $\gamma\gamma$ and $Z \gamma$ decay modes, which lead to 
clear signals, are  very rare with rates of ${\cal O}(10^{-3})$. Note that for
Higgs masses around 135 GeV, the decay $H\to WW^* \to W f\bar f$ although at
the  three--body level starts to dominate over the two--body $H\to b\bar b$
mode: the much larger $HWW$ coupling compared to $Hb\bar b$ compensates for the
suppression by  the additional electroweak coupling and the virtuality of the
$W$ boson. 

In the ``high--mass" range, $M_H \gsim 140$ GeV, the Higgs bosons decay into
$WW$ and $ZZ$ pairs, one of the gauge bosons being possibly virtual  below the
thresholds. Above the $ZZ$ threshold, the branching ratios are 2/3 for $WW$ and 
1/3 for $ZZ$ decays, and the opening of the $t\bar{t}$ channel for higher $M_H$
does not alter  this pattern significantly.  

In the low--mass range, the Higgs is very narrow, with $\Gamma_H<10$ MeV, but
this width increases, reaching 1 GeV at the $ZZ$ threshold. For  very large
masses, the Higgs  becomes obese, since $\Gamma_H \sim M_H$, and can hardly be
considered as a resonance. 

\begin{figure}[!h]
\vspace*{-1.5cm}
\hspace*{-1.8cm}
\begin{tabular}{ll}
\begin{minipage}{10.cm} 
\hspace*{-.4cm}
\centerline{
\includegraphics[scale=0.33]{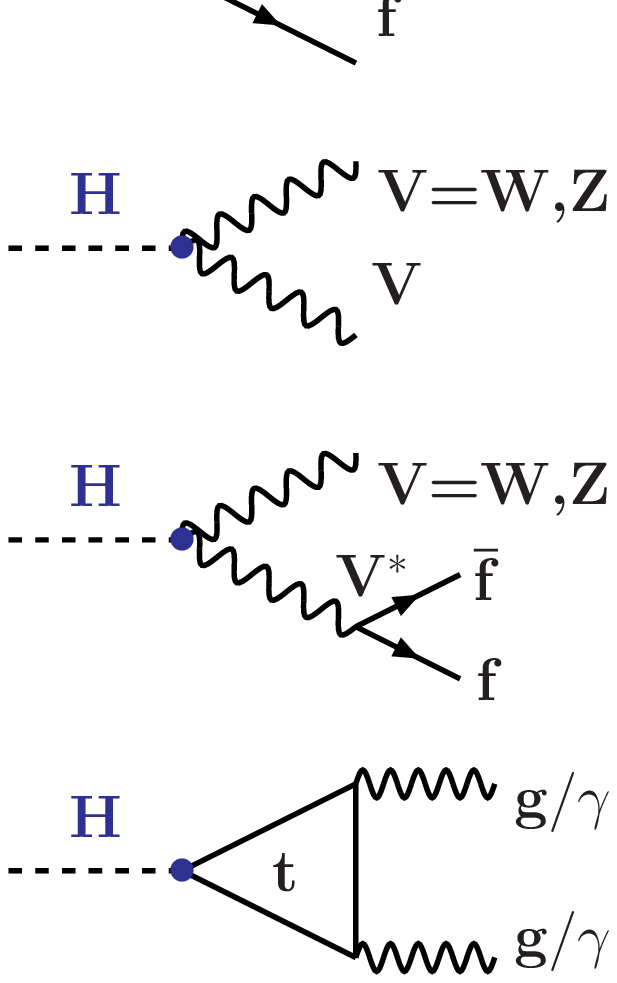}\hspace*{-4.2cm}
\includegraphics[scale=0.44]{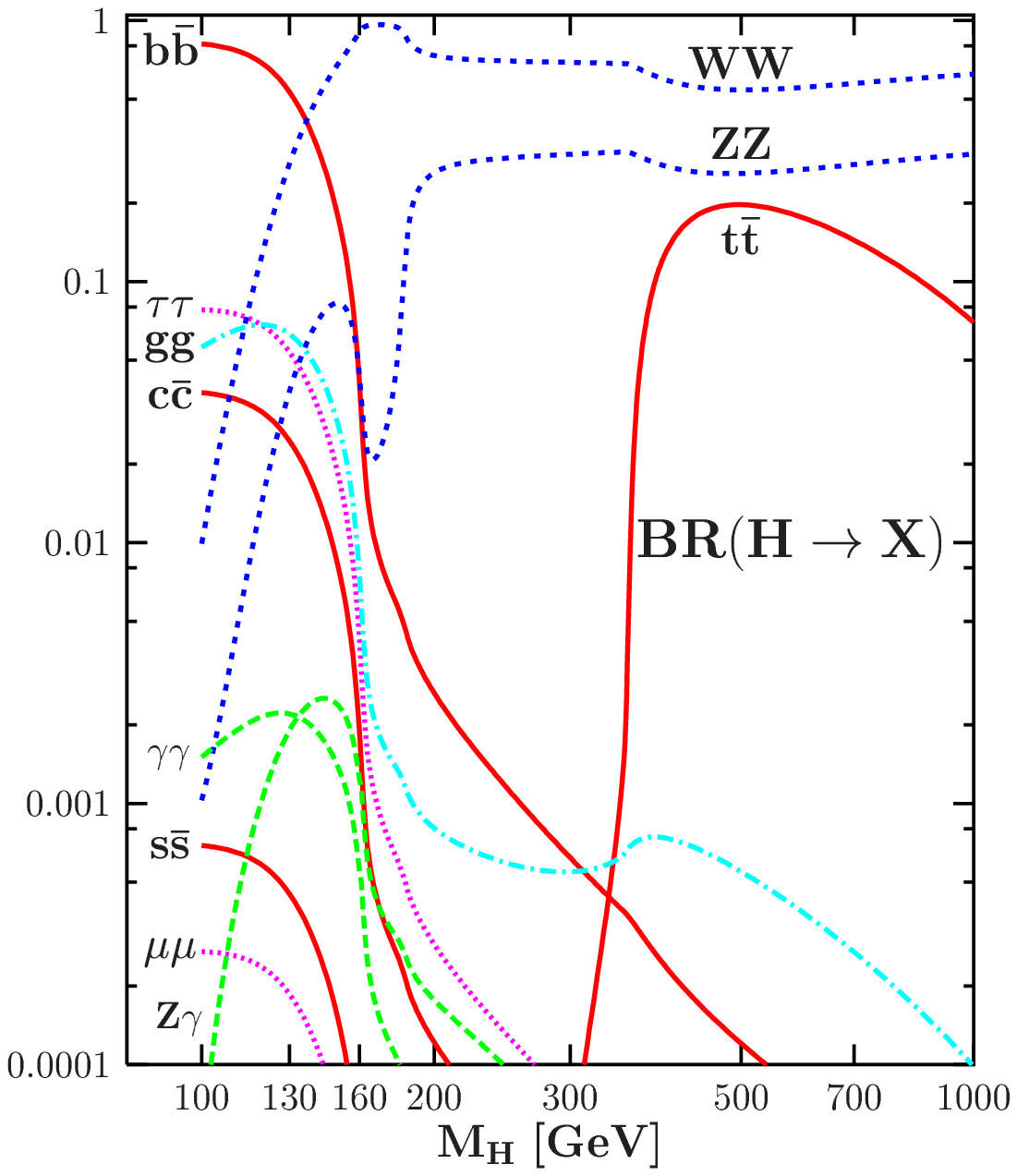} 
}
\end{minipage}
& \hspace*{-2.cm}
\begin{minipage}{5cm}
\vspace*{-1cm}
\includegraphics[width=5cm,height=10cm]{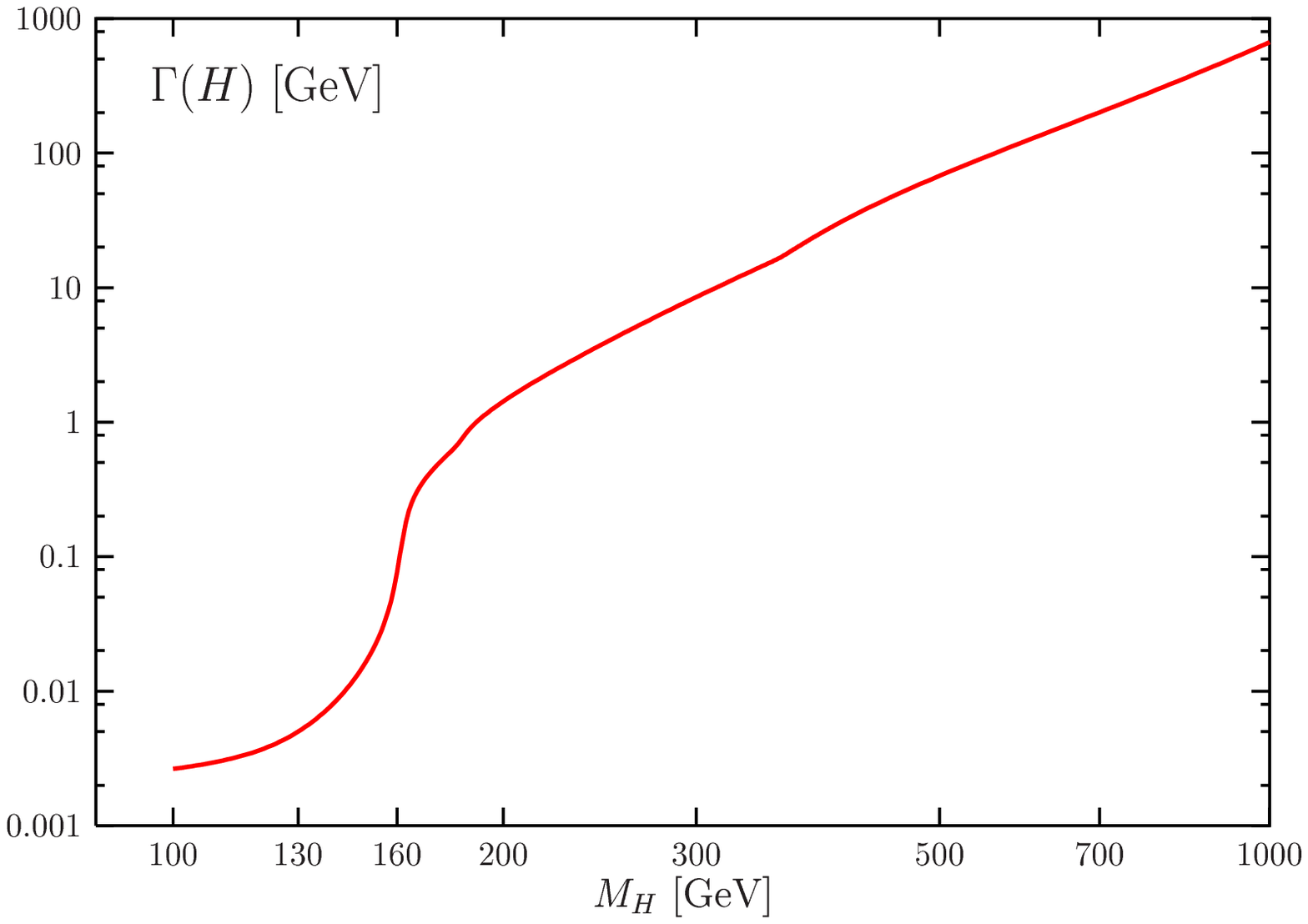}\vspace*{-16.6mm}
\end{minipage}
\vspace*{-6.cm}
\end{tabular}
\caption{The decays of the SM boson: Feynman diagrams (left), branching ratios 
(center) and the total decay width (right) as a function of its mass; from 
Ref.~\cite{hdecay}.} 
\label{Hfig:brwidsm}
\vspace*{-.3cm}
\end{figure}

The branching ratios and total decay widths are summarized in
Fig.~\ref{Hfig:brwidsm}, which is  obtained from a  recently updated version of
the Fortran code {\tt HDECAY} \cite{hdecay} which includes all relevant channels
with the important radiative corrections and other higher order effects 
\cite{decays}.  In addition, the theoretical uncertainties on the Higgs
branching ratios  should also be considered. Indeed, while the Higgs decays into
lepton and gauge boson pairs are well under control (as mainly small electroweak
effects are involved),  the partial decays widths into quark pairs and gluons
are plagued with  uncertainties  that are mainly due to  the imperfect knowledge
of the bottom and charm  quark masses and the value of the strong coupling
constant $\alpha_s$.  At least in the intermediate mass range, $M_H \approx
120$--150 GeV where the SM Higgs decay rates into $b\bar b$ and $W^+W^-$ final
states have the same order of magnitude, the parametric uncertainties on these
two main Higgs decay branching ratios are non--negligible, being of the order of
3 to 10\% \cite{pap-lhc,decays-errors}.

\subsection{Detection channels at the Tevatron and the LHC} 

From the previous discussion, one concludes that producing the SM Higgs particle
is relatively easy; this is particularly the case at the LHC  thanks to the high
energy of the collider and its expected  luminosity. However, detecting the
particle in a very complex hadronic environment is another story. Indeed,   in
the main Higgs decay channels the backgrounds  are simply gigantic
\cite{backgrounds}.  For instance, the rate for the production of light quarks
and gluons is ten orders of magnitude larger than that of the Higgs boson.  Even
the cross sections for the production of $W$ and $Z$ bosons are three to four
orders of magnitude larger. Detecting the Higgs particle is this hostile
environment resembles to finding a needle in a (million) haystack, the challenges
to be  met being simply enormous. 

To be able to detect the Higgs particle, one should take advantage in an optimal
manner of the kinematical characteristics of the signal events which are, in
general, quite different from that of the background events. In addition, one 
should focus on the decay modes of the Higgs particles (and those of the 
particles that are produced in association such as $W^\pm,Z$ bosons or top
quarks) that are easier to extract from the background events. Pure hadronic
modes such as Higgs decays into light quark or gluon jets have to  be discarded
although much more frequent  in most cases. 

\begin{figure}[hbtp]
\vspace*{-3mm}
\centerline{
\epsfig{file=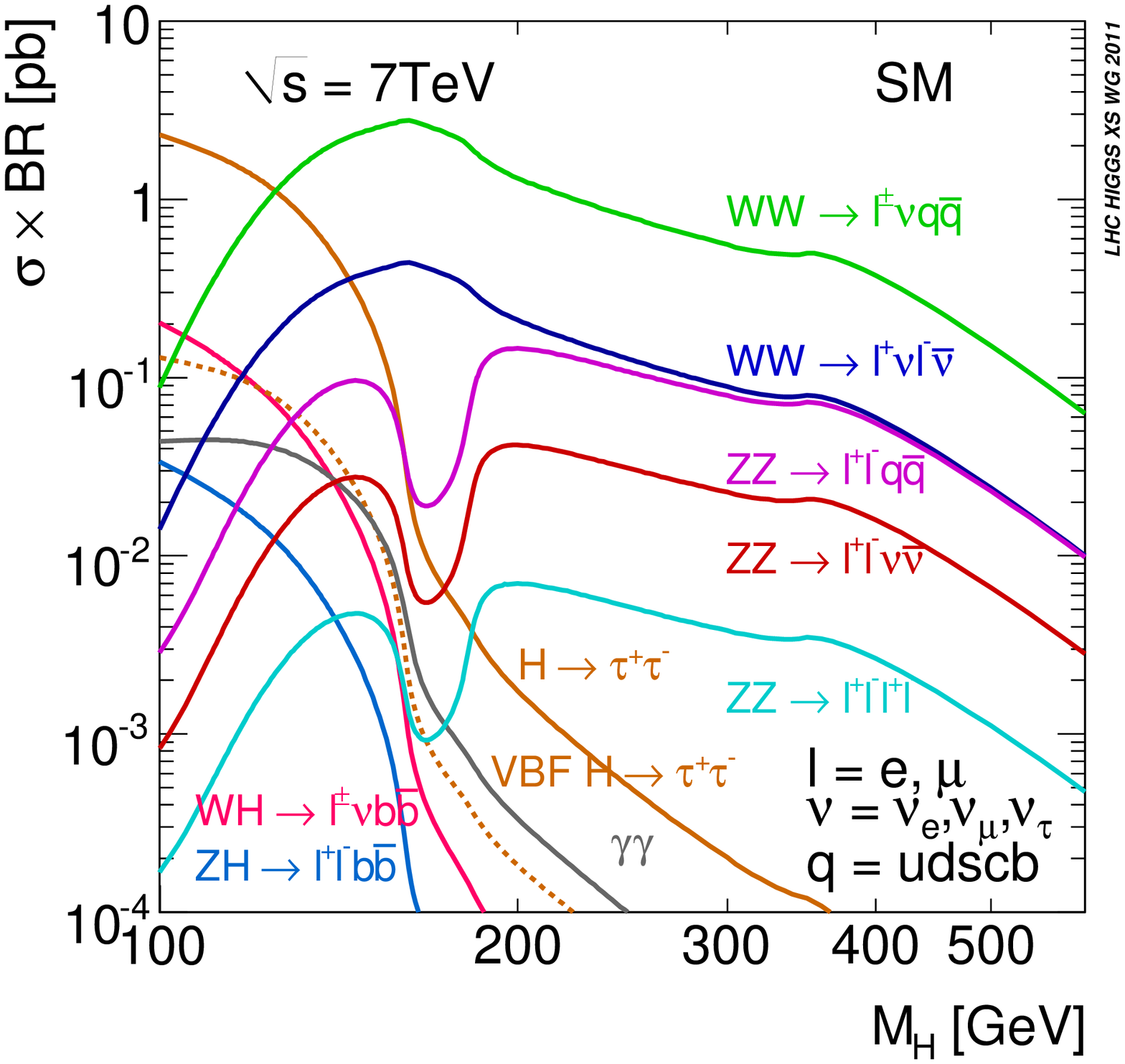,width=4.8cm}
\epsfig{file=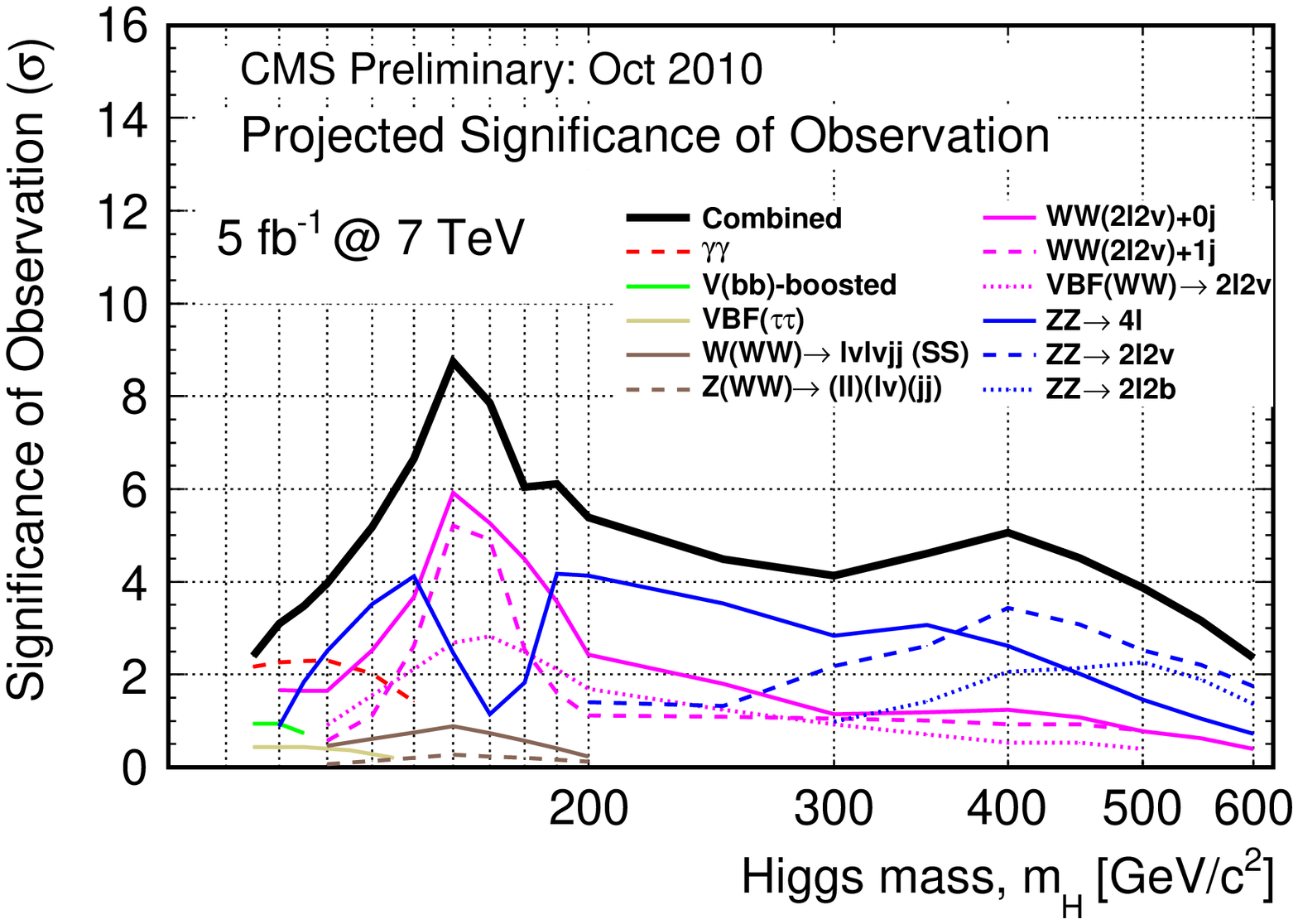,width=6cm}
}
\vspace*{-4mm}
\caption{Left: Higgs cross sections times branching ratios for interesting 
final states \cite{LHCXS}. Right: Projected significance for the observation 
of the SM Higgs in various channels at the LHC with $\sqrt s=7$ TeV and 
5 fb$^{-1}$ data in the CMS experiment \cite{Sharma}.}
\label{sensitivity} 
\vspace*{-2mm}
\end{figure}

The Higgs detection channels  have been discussed in great details at this
conference \cite{Tevatron,Nisati,Sharma} and I will simply recall a few basics
elements. At the LHC with $\sqrt s=7$ TeV, the most  promising channels, with
rates as shown in Fig.~\ref{sensitivity} (left), are as follows.

In the $gg$ fusion mechanism, the detection channels for a light Higgs boson, 
$M_H \lsim 160$ GeV are \cite{gg-detection}:  $H \to \gamma \gamma$ (mostly for
$M_H \lsim 140$ GeV),  $H\to ZZ^* \to 4\ell^\pm$ and $H\to WW^{(*)}\to \ell \ell
\nu \nu$ with $\ell=e,\mu$ (for masses below, respectively, $2M_W$ and $2M_Z$).
For $160 \lsim M_H \lsim 180$ GeV, only $H \to WW \to \ell \ell \nu \nu$ is 
possible. For higher masses, $M_H \gsim 2M_Z$, it is the golden mode $H\! \to\!
ZZ\! \to\! 4\ell^\pm$, which for slightly higher $M_H$  can be supplemented by $H
\! \to \! ZZ \! \to \ell^+ \ell^-  \nu\bar{\nu}, \ell^+ \ell^- jj$ and $H \!
\to\!  WW \! \to \!  \ell \nu jj$ to increase the statistics.  Recently, the
inclusive channel $gg\! \to\! H\!  \to\! \tau^+ \tau^-$ used in the MSSM
appeared to be also feasible in the SM  for Higgs masses below $\approx 140$
GeV  \cite{htau}. Most of these channels could a allow a 2$\sigma$ sensitivity
on the Higgs boson with  5--10  fb$^{-1}$ data, as shown in the right-hand side
of Fig.~\ref{sensitivity}.  Note that while the signal peaks are very narrow in
the  $H\to \gamma \gamma, 4\ell$ cases (with a $\approx 1$ GeV resolution in the
low  mass range when the Higgs total width is very small), they are much
wider in the $H\to \ell \ell \nu \nu$ and $\tau^+ \tau^-$ channels.  

The signal sensitivity, in particular in $H\!\to\! WW^*, \tau\tau$, can be
improved by considering jet categories, where the total $gg\!\to\! H$ cross
section is broken into Higgs plus 0, 1 and 2 jet cross sections which are known
at NNLO, NLO and LO, respectively; see for instance Ref.~\cite{ggH-ADGSW}. One
can significantly reduce  the backgrounds, in particular in the $H$+0j and
$H$+1j categories  which are little affected by the $t\bar t$ background.
However, as pointed out in Ref.~\cite{Scale-H0j}, the scale uncertainty for the
separate rates will  increase to the 20\% level. 

Vector boson fusion will lead to interesting signals at the LHC but presumably
only at the highest energies. This process is of particular interest since, as
discuss previously,   it has a large enough cross section  which is affected
only little by theoretical uncertainties. In addition, one can use specific
cuts  (forward--jet tagging, mini--jet veto for low luminosity as well as
triggering on the central Higgs decay products) \cite{WWfusion0}, which render
the backgrounds comparable to the signal, therefore allowing precision Higgs
coupling measurements. It  has been shown in parton level analyses (as well as
detailed simulations for some channels) that the decay $H \to \tau^+ \tau^-,
WW^*$ and  possibly $H \to \gamma \gamma , ZZ^*, WW^*$ can be detected
\cite{WWfusion} and could allow for coupling measurements.

The associated production with gauge bosons, $q\bar q \to HV$  with $H \to
b\bar{b}$ and possibly $H \to WW^* \to \ell \ell \nu \nu $ and $\ell \nu  jj$,
is the most relevant detection mechanism at the Tevatron for a light Higgs boson
\cite{Tevatron}, $gg \to H \to W W \to \ell \nu \ell \nu$ being important for
Higgs masses  around 160 GeV. At the LHC, this process was expected to play
only a marginal role with eventually   only the process $HW\! \to\! \ell \nu
\gamma \gamma$  observable at sufficiency high energy and with a large amount of
luminosity  \cite{VH-Kunszt}. Recently, however, it has been shown that in
$q\bar q\! \to\!  HV$ with $H\! \to\! b\bar b$, the statistical significance of the
signal can be significantly increased by looking at boosted jets when the Higgs
has large transverse momenta \cite{VH-boost}. This channel can be thus
used with sufficient data and,  as it is theoretically clean, it would allow for
measurements of the Higgs couplings to gauge bosons  and bottom quarks.

Finally, associated $t\bar t H$ production with $H \to \gamma \gamma$
\cite{ttH-gamma} or $b\bar{b}$ \cite{ttH-bb}, can in principle be observed at
the LHC and allows direct measurement of the top Yukawa coupling.  The $H\to
b\bar b$ channel would also allow to access the bottom Yukawa coupling and,
eventually, a determination of the Higgs CP properties.   Unfortunately, 
detailed analyses have  shown that $pp \to t\bar t H\to t\bar t b\bar b$ might
be subject to a too large jet background in addition to the irreducible $t\bar t
b\bar b$ background \cite{ttH-bkg}. Nevertheless, the recently  advocated 
boosted jet techniques together with more refined  analyses might resurrect this
channel \cite{ttH-boost}.

In Fig.~\ref{sensitivity}, the sensitivity of some of the channels discussed
above is shown at the LHC with $\sqrt s=7$ TeV c.m. energy and 5 fb$^{-1}$ 
integrated luminosity per experiment.  It is clear that we are reaching the
stage  at which the Higgs particle should be observed. More details are given in
the CDF/D0, ATLAS and CMS talks \cite{Tevatron,Nisati,Sharma}.

\section{The Higgs particles in supersymmetric theories}

\subsection{The Higgs sector of  the MSSM} 

In supersymmetric extensions of the SM \cite{MSSMbook}, at least two Higgs
doublet fields are required for a consistent electroweak symmetry breaking and
in the minimal model, the MSSM,  the Higgs sector is extended to contain five
Higgs bosons: two CP--even $h$ and $H$, a CP-odd $A$ and two charged Higgs
$H^\pm$ particles \cite{HHG,Review2,Sven}.  Besides the  four masses,  two more
parameters enter the MSSM Higgs sector: a mixing angle $\alpha$ in the neutral
CP--even sector  and the ratio of the  vacuum expectation values of the two
Higgs fields $\tb$. In fact,  only two free parameters are needed at
tree--level: one Higgs mass, usually chosen to be $M_A$ and   $\tan\beta$ which 
is expected to lie in the range $1 \lsim \tb \lsim m_t/m_b$.  In addition, 
while the  masses of the heavy neutral and charged $H,A,H^\pm$ particles are
expected to range from $M_Z$ to the SUSY breaking scale $M_S={\cal O}(1$ TeV), 
the mass of the lightest Higgs boson $h$ is bounded from above, $M_h \le M_Z$ at
tree--level.  This relation is altered by large radiative corrections,  the
leading part of which grow as the fourth power of $m_t$ and logarithmically with
the SUSY scale  or common squark mass $M_S$; the mixing (or trilinear coupling)
in the stop sector $A_t$ plays also an important role.  The upper bound on $M_h$
is then shifted to $M_h^{\rm max} \sim 110$--135 GeV depending on these
parameters \cite{Sven}.

For a heavy enough  $A$ boson, $M_A \gg M_Z$,  $h$  reaches its  maximal mass
value $M_h\! \simeq\!  M_h^{\rm max}$ and has SM--like couplings to fermions
and gauge bosons. In this decoupling regime \cite{decoupling}, the three other
Higgs bosons are  almost degenerate in mass, $M_H \!\approx\! M_A\! \approx\!
M_H^{\pm}$ and  the couplings of the CP--even $H$ boson (as well as those of the
charged $H^\pm$  bosons) become similar to that of the $A$ boson: no tree--level
couplings to the gauge bosons and couplings to isospin down (up) type fermions 
that are (inversely) proportional to $\tb$. In particular, for high  $\tb \gsim
10$ values, the $H, A$ Yukawa couplings to $b$--quarks and $\tau$--leptons are
strongly  enhanced and those to $t$--quarks strongly suppressed. 

For a light pseudoscalar boson,  $M_A \lsim M_h^{\rm max}$ at high $\tb$,  one
is in the antidecoupling regime \cite{antidecoupling} in which  the roles of the
CP--even $h$ and $H$ states are reversed: it is the $H$ boson which has a mass
$M_H\simeq  M_h^{\rm max}$ and SM--like couplings, while  the $h$ particle
behaves exactly like the pseudoscalar $A$ state, i.e.  $M_h \simeq M_A$, no
couplings to gauge bosons and enhanced (suppressed) ones to $b, \tau$ ($t$)
states.  

We are thus always in a scenario where one has a SM--like state $H_{\rm SM}\!=
\!h(H)$ and two CP--odd like Higgs particles $\Phi\!=\!A$ and $H(h)$ when we are
in  the decoupling (anti\-decoupling) regime which, for $\tb \! \gsim \! 10$,
occurs  already for $M_A \! \gsim \! M_h^{\rm max}$  ($M_A \!\lsim \!M_h^{\rm
max}$).

There is an intermediate scenario for $\tan \beta \gsim 10$: the intense
coupling regime \cite{intense} in which the three neutral states have comparable
masses, $M_h \approx M_H \approx M_A \approx M_h^{\rm max}$ and couplings to
isospin down type fermions that are  enhanced; the squares of the CP--even Higgs
couplings approximately add to the square of the CP--odd Higgs coupling. 

Finally, there is the regime in which the superparticles are light enough to
affect Higgs phenomenology. Higgs bosons can decay into charginos, neutralinos
and even  sfermions and they can be produced in the decays of these sparticles.
Sparticle loops could also alter the Higgs production and decay pattern. These
``dream" scenarios in which both  Higgs and sparticles are accessible will not
be addressed; see Ref.~\cite{Review2} for discussions.

In the most general case, the decay pattern of the MSSM Higgs particles can be
rather complicated, in particular for the heavy states. Indeed, besides the
standard decays into pairs of fermions and gauge bosons, the latter can have
mixed  decays into gauge and Higgs bosons while the $H$ boson can decay into
$hh$ states. However,  for the large values of $\tb$ that are interesting for
the  Tevatron and the early  LHC, $\tb \gsim 10$, the couplings of the non--SM
like Higgs bosons to $b$ quarks and $\tau$ leptons are so strongly enhanced and
those to top quarks and gauge bosons suppressed, that the pattern becomes very
simple. To a very good approximation, the $\Phi=A$ or $H(h)$  bosons   decay 
almost exclusively into $b\bar b$ and $\tau^+\tau^-$ pairs,  with branching
ratios of, respectively, $\approx 90\%$ and $ \approx 10\%$, while the $t\bar t$
channel and the decays involving gauge or Higgs bosons are suppressed to a level
where the branching ratios are less than 1\%. The CP--even  $h$ or $H$ boson,
depending on whether we are in the decoupling or antidecoupling regime, will
have the same decays as the SM Higgs boson  in the mass range below $M_{H_{\rm
SM}}^{\rm max} \lsim 135$ GeV. Finally,  the $H^\pm$ particles decay into
fermions pairs: mainly $t\bar{b}$ and $\tau \nu_{\tau}$ final states for $H^\pm$
masses, respectively, above and below the $tb$ threshold. Adding up the various
decays, the widths of all five Higgses remain  rather narrow.  For a detailed
discussion of MSSM Higgs decays, see Ref.~\cite{Review2}.\vspace*{-2mm}

\subsection{The production cross sections at the LHC} 

In the MSSM, the dominant production processes for the CP--even neutral $h$ and
$H$ bosons are essentially  the same as those for the SM Higgs particle
discussed before. In fact, for $h (H)$  in the decoupling (antidecoupling)
regimes, the cross sections are almost exactly the same as for  the SM Higgs
particle with a mass $\approx M_h^{\rm max}$. In the case of the pseudoscalar
$A$ boson, the situation is completely different. Because of CP invariance which
forbids tree--level $A$ couplings to gauge bosons, the $A$ boson cannot be
produced in the Higgs-strahlung and vector boson fusion processes;  only the
$gg\to A$ fusion  as well as associated production with  heavy quark pairs,
$q\bar q, gg \to Q\bar Q A$, will be in practice relevant (additional processes,
such as associated production of CP--even and CP--odd Higgs particles, have too 
small cross sections). This will therefore be also the case of the CP--odd  like
particles $\Phi=H(h)$ particles in the decoupling (antidecoupling)  regime.

If one concentrates on the moderate to high $\tb$ regime that is the relevant
one at both the Tevatron and the early LHC with $\sqrt s =7$ TeV,  the
$b$--quark will play a major  role: as its couplings to the CP--odd like 
$\Phi=A$ or $H(h)$ states are strongly enhanced, only processes involving the
$b$--quark will be important.  Thus, in the $gg\!  \to\!  \Phi$  processes, 
one  should take into account  the $b$--quark loop  which provides the dominant
contribution  and in associated Higgs production with heavy quarks, $b\bar{b}$
final states must be considered.  

In the $gg \to \Phi$ processes with only the $b$--quark loop included,   as
$M_\Phi \gg m_b$, chiral symmetry approximately holds and the cross sections
are approximately the same for the CP--even $H\;(h)$ and CP--odd $A$ bosons. The
QCD corrections are known only to  NLO for which the exact calculation with
finite loop quark masses is available \cite{SDGZ}. Contrary to the SM case, 
they increase only moderately the production cross sections. 

In the case of the $pp \to b\bar b \Phi$ processes, the NLO QCD corrections have
been calculated in Ref.~\cite{gg-bbH-QCD} and turn out to be rather large.  
Because of the small $m_b$ value, the cross sections  develop large logarithms
$\log(Q^2/m_b^2)$ with the scale $Q$ being typically of the order of the
factorization scale $\mu_F\! \sim\! M_\Phi\! \gg\! m_b$. These can be  resummed
by considering the $b$--quark as a massless parton and  using heavy quark
distribution functions at a scale  $\mu_F\! \sim\! Q$ in a five active flavor
scheme. In this scheme, the inclusive process where one does not require to
observe the $b$ quarks is simply the $2\! \to\! 1$ process $b \bar b\! \to\!
\Phi$ at LO \cite{bbH-LO}. If one requires the observation of one  high--$p_T$
$b$--quark, one has to consider  the NLO  corrections \cite{bbH-NLO} and in
particular the $2\!\to\! 2$ process $gb\! \to\! \Phi b$. Requiring the
observation of two $b$ quarks,  one has to consider the $2\! \to\! 3$ process
$gg \to b\bar b \Phi$ discussed previously, which is the leading mechanism at
NNLO \cite{bbH-NNLO}. Thus, instead of  $q\bar q, gg\! \to\! b\bar b \Phi$, one
can consider the process $b\bar b \to \Phi$ for which  the cross section is
known up to NNLO in QCD \cite{bbH-NLO,bbH-NNLO}, with corrections that are of
moderate size if $i$) the bottom quark mass in the Yukawa coupling is defined at
the scale $M_\Phi$ to absorb large logarithms $\log(\mu_R^2/m_b^2)$ and $ii)$ 
if the  factorization  scale is chosen to be small, $\mu_F=\mu_R=\mu_0= \frac14
M_\Phi$.

\begin{figure}[!h] 
\begin{center} 
\mbox{\hspace*{-3mm}
\epsfig{file=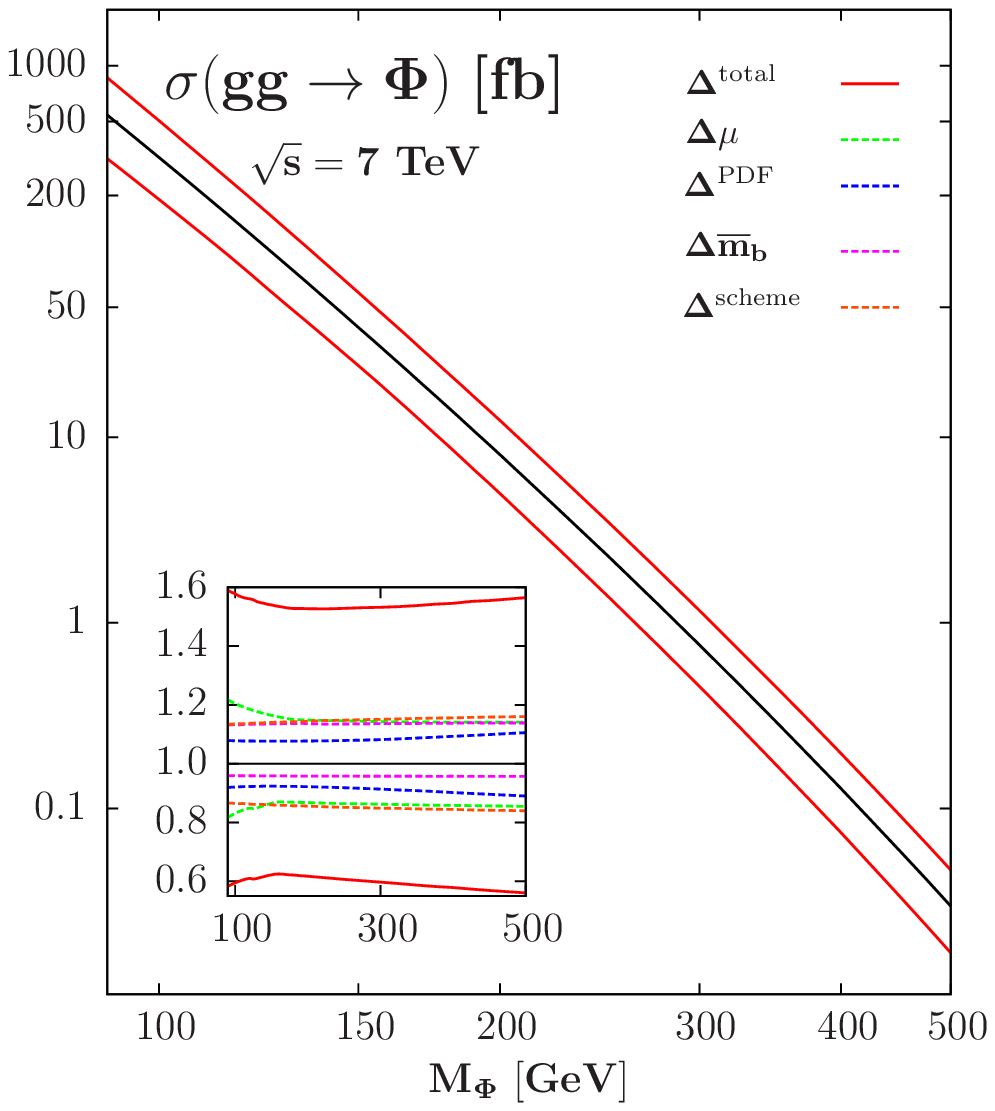,width=4.3cm}\hspace*{-.2mm}
\epsfig{file=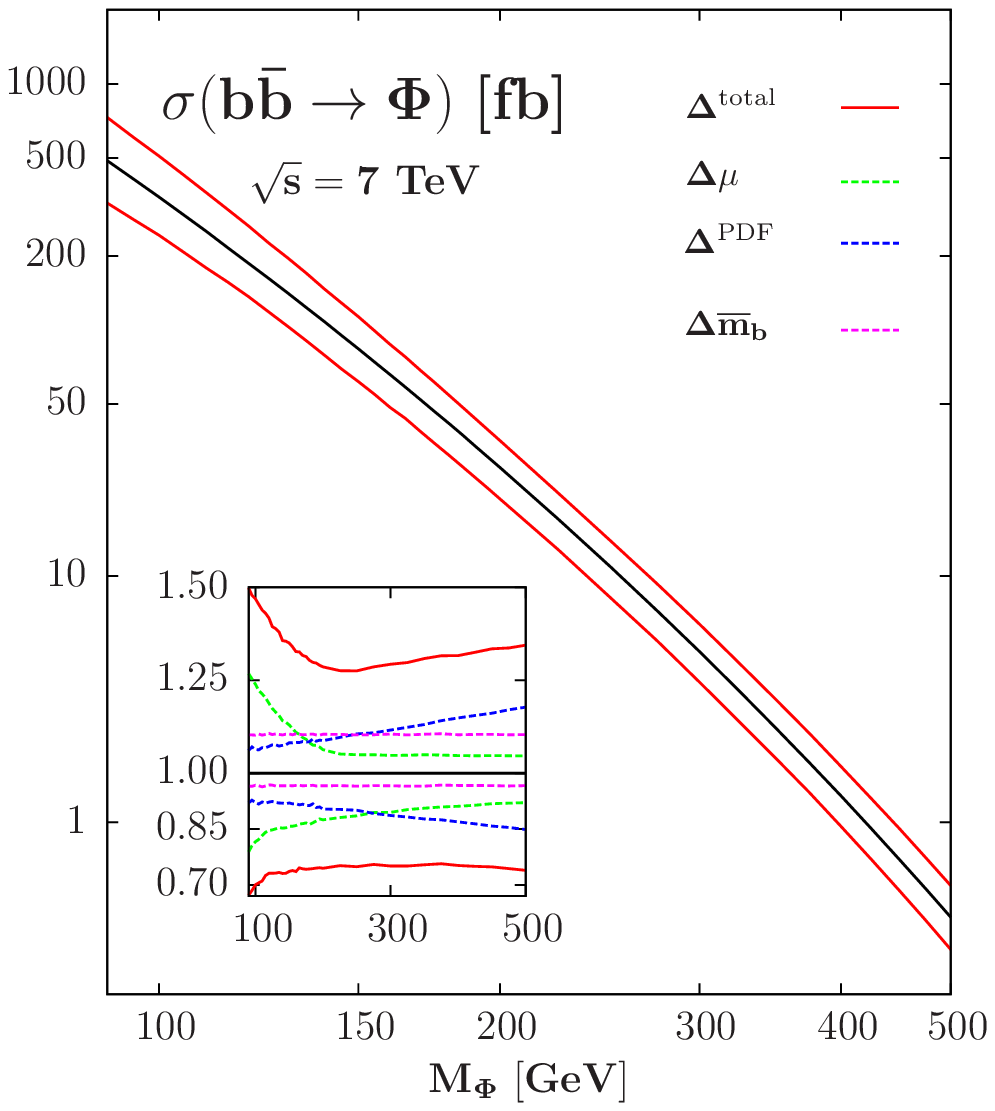,width=4.3cm}\hspace*{-.2mm}   
\epsfig{file=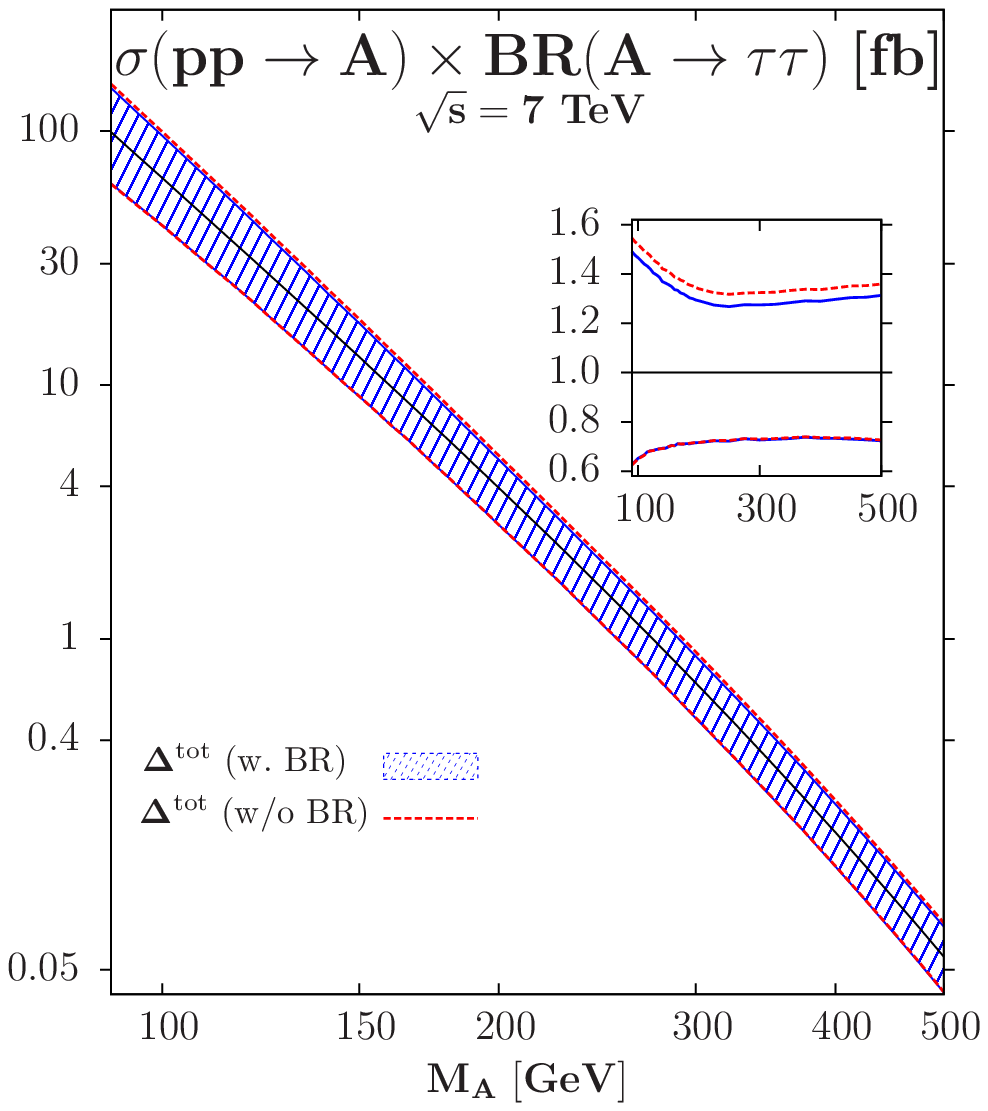,width=4.3cm} }
\end{center} 
\vspace*{-5mm}
\caption[]{The cross sections $\sigma^{\rm NLO}_{gg\!\to \!A}$ (left) and 
$\sigma^{\rm NNLO}_{b\bar b \!\to\!A}$ (center) at LHC energies as a function 
of $M_A$ when using the MSTW PDFs and unit $Ab\bar b$ couplings (the true cross
section are obtained by multiplying by 2$\tan^2\beta$)  and the various
individual and total uncertainties (scale, PDF and those due to $m_b$). The 
combined $\sigma(p p\! \to\! A)\times  {\rm BR}(A \!\to\! \tau^+ \tau^-)$ rate
and total uncertainties with and  without the branching ratio is shown in the
right panel. In the inserts, shown  are the various theoretical uncertainties
when the rates are normalized to   the central values. From \cite{pap-lhc}.}
\vspace*{-1mm}
\label{pp-htau}
\end{figure}

The cross sections for these two processes have been updated in 
Refs.~\cite{pap-lhc,LHCXS} for the LHC. We will not discuss the details here
and  simply summarize the main results in Fig.~\ref{pp-htau} where the cross
sections, together with the associated theoretical uncertainties, are
shown.\vspace*{-3mm}

\subsection{Sensitivity at the LHC}

\begin{figure}[!h]
\begin{center}
\vspace*{-.2mm}
\epsfig{file=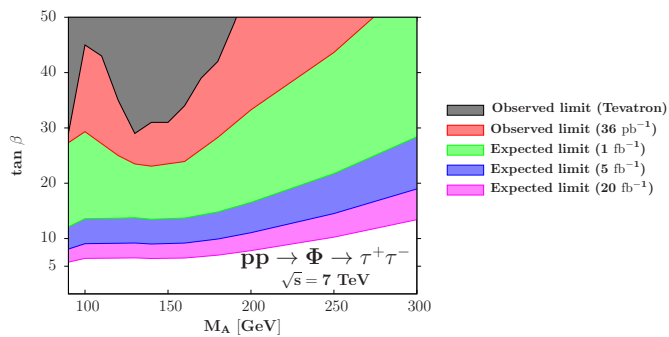,scale=0.5} 
\end{center}
\vspace*{-5mm}
\caption[]{Contours for the expected $\sigma(p p\! \to\! \Phi \!\to 
\!\tau^+ \tau^-)$ 95\% CL exclusion limits at the LHC with $\sqrt s\!=\!7$ TeV 
in  the [$M_A, \tb$] plane for various luminosities. The ``observed" limits 
from CMS with 36 pb$^{-1}$ data and Tevatron are also displayed \cite{htau}. } 
\vspace*{-3mm}
\label{projection}
\end{figure}

We now summarise  the  ATLAS and CMS sensitivity on the MSSM parameters  in the
important $p p \! \to \! \Phi \! \to \! \tau^+ \tau^-$ with inclusive channel
which combines $gg\! \to \! \Phi$ and $b\bar b\! \to \Phi$. We consider the
``observed''  or ``expected" values of the cross section times branching ratio
that have been given by the CMS  collaborations  with  36 pb$^{-1}$ data for the
various values of $M_A$ and turn them into exclusion limits in this plane by
simply rescaling $\sigma(gg+b\bar b\! \to\! A \to \tau\tau)$ of 
Fig.~\ref{pp-htau} by a factor $2\times\! \tan^2 \beta$.  This is shown in
Fig.~\ref{projection} where  the [$\tb, M_A$] exclusions contours from the LHC
the   Tevatron are presented. 

One observes that the CMS exclusion limits on the $[\tb, M_A$] MSSM
parameter space with  only  36 pb$^{-1}$ data are extremely strong as, for
instance, values $\tb \gsim 30$ are excluded in the low $A$ mass range,
$M_A=90$--200 GeV.  If the luminosity is increased to the fb$^{-1}$ level, one
obtains constraints that are similar to those presented at this conference by
ATLAS and CMS \cite{Nisati,Sharma}.  Assuming that there will be no improvement
in the analysis (which might be a little pessimistic) and that the CMS
sensitivity will simply scale as the square root of the integrated luminosity,
the region of the $[\tb,M_A]$ parameter space which can be excluded in the case
where no signal is observed is also displayed in  Fig.~\ref{projection} for
several values of the accumulated luminosity. With 5  fb$^{-1}$ data per
experiment (or with half of these data  when the ATLAS and CMS results are
combined), values $\tb \gsim 9$ could be excluded in the mass range around $M_A
\approx 130$ GeV. The search is sensitive to values $\tb \approx 6$ at
$M_A=115$--140 GeV with 20 fb$^{-1}$ data.

We should note that these exclusions limits have been obtained in the so--called
maximal mixing scenario which is used as benchmark for MSSM Higgs studies 
\cite{S-benchmark}. However they equally hold in other mixing scenarios (such as
no--mixing) as the only difference originates from the potentially large
SUSY-corrections  to the Higgs--$b\bar b$  Yukawa couplings, which
cancel out in the cross section times branching ratio; see 
Ref.~\cite{pap-lhc,htau}.

These very strong limits on the MSSM parameter space from $pp\!\to\! \Phi \!\to
\! \tau\tau$  could be further improved by considering four additional
production channels.

-- The process  $gb \to \Phi b \to b b\bar b$ where the final bottom quarks are
detected: the production cross section  is one order of magnitude lower than
that of the inclusive $gg+b\bar b\to \Phi$ process  but this is compensated by
the larger fraction BR$(\Phi \to b\bar b) \approx 90\%$ compared to BR$(\Phi \to
\tau^+ \tau^-) \approx 10\%$; the QCD background are much larger though.

-- The process  $pp \to \Phi \to \mu^+\mu^-$ for which the rate is simply  
$\sigma(pp\!\to\! \Phi \to \tau\tau )$ rescaled by BR$(\Phi \to \mu
\mu)/$BR$(\Phi \to \tau \tau)= m_\mu^2/m_\tau^2 \approx 4\times 10^{-3}$; the
smallness of the rate is partly compensated by the much cleaner $\mu \mu$ final
state and the better resolution on the $\mu\mu$ invariant mass.  In particular, 
this small resolution could allow to separate the three peaks of the almost
degenerate $h,H$ and $A$ states in the intense coupling regime
\cite{intense}.

-- The process $pp \to tbH^- \to tb \tau \nu$ which leads to a cross section
that is also proportional to $\tan^2\beta$ (and which might also be useful for
very low $\tb$  values) but that is two orders of magnitude smaller  than
$\sigma(pp \to \Phi)$ for $M_A \approx 100$--300  GeV. 

-- Charged Higgs production from top quark decays, $pp\! \to\! t\bar t$ with 
$t\to H^+b \to \tau^+ \nu b$, which has also been recently analyzed by the 
ATLAS and CMS collaboration \cite{Sharma}. This channel should ultimately cover
the range $M_{H^\pm}\! \lsim \! 160$ GeV  independently of $\tb$.

Unfortunately, in the four cases, the small rates  will allow only for a modest
improvement over the $pp \to \Phi \to \tau \tau$ signal or exclusion limits. In
fact, according to the (presumably by now outdated) projections of the ATLAS and
CMS collaborations \cite{atlastdr,CMSTDR}  at the full LHC with $\sqrt s=14$ TeV
and 30 fb$^{-1}$ data, these processes are observable only for  not too large
values of $M_A$  and relatively high values of $\tan\beta$ ($\tb \gsim 20$) most
of which are already  excluded. Nevertheless, as is the case for the $pp\to \tau
\tau$ channel, some (hopefully significant) improvement over these projections
might be achieved and a better sensitivity reached at the 14 TeV LHC with 
$\gsim 100$ fb$^{-1}$ data.\vspace*{-2mm}

\section{Measurements of the Higgs properties}

\begin{figure}[hbtp]
\begin{center}
\vspace*{-1.4cm}
\epsfig{file=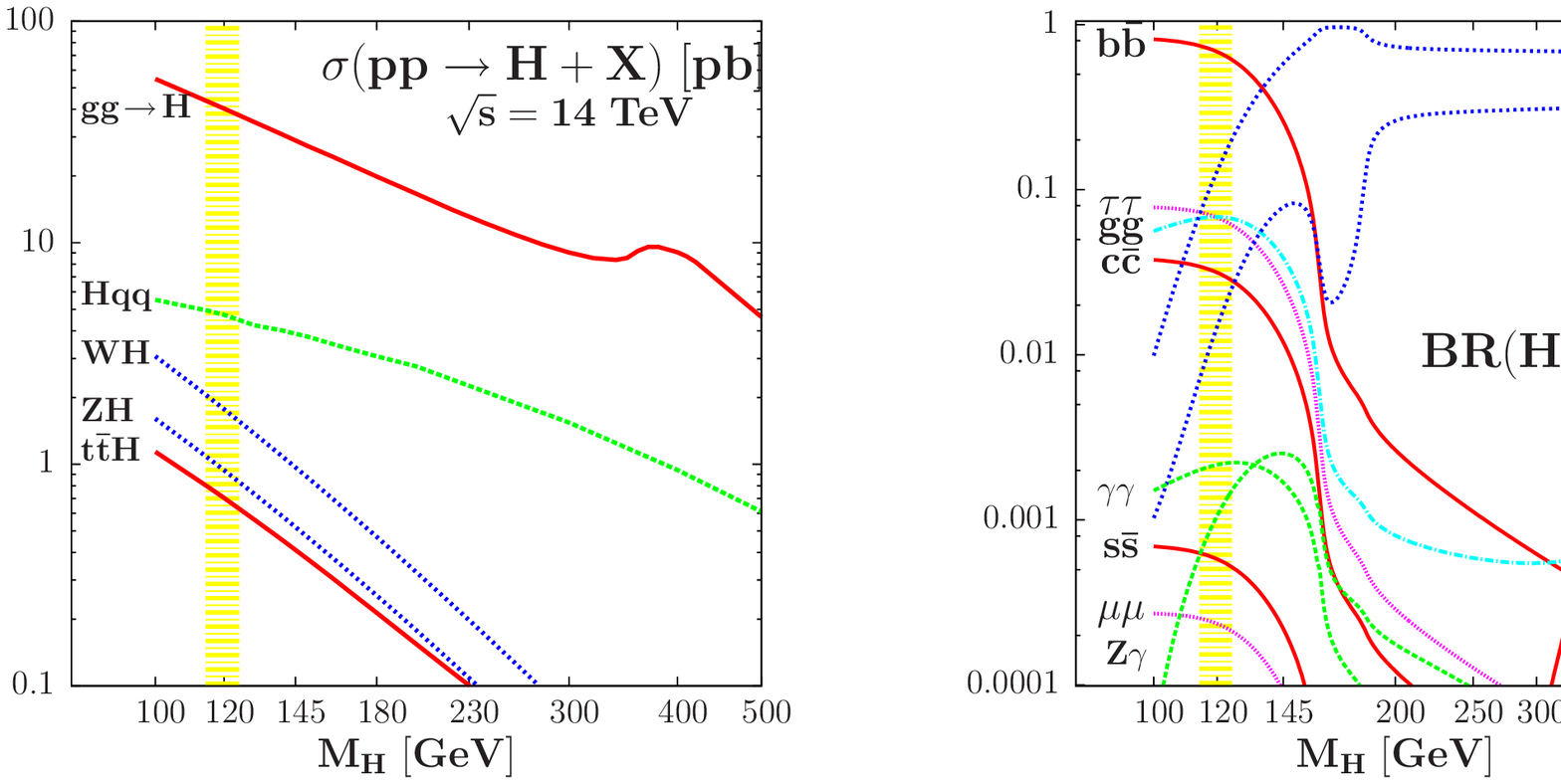,width=8cm}
\vspace*{-6.3cm}
\hspace*{-1cm}
\caption[]{The Higgs cross sections at the 14 TeV LHC in the various channels
(left) and the Higgs branching ratios (right); the range $115\lsim M_H
\lsim 130$ GeV is highlighted.}
\end{center}
\label{allH}
\vspace*{-.9cm}
\end{figure}

Observing the Higgs particles is only the fist part of our contract, as  it is
of equal importance that, after seeing the Higgs signal at the LHC, we perform
measurements of the Higgs properties, to be able to establish the exact nature
of EWSB and to achieve a more fundamental understanding of the issue.   In this
section, we address the Higgs properties  question  at the LHC when the maximal
energy $\sqrt s\sim 14$ TeV  has been  reached and a large  luminosity, $\gsim
100$ fb$^{-1}$, has been collected. In fact, it appears that for a light  Higgs
boson with a mass below 130 GeV, many interesting measurements could be
performed at the LHC as for such masses: $a)$ the Higgs boson is accessible in
all the main production channels  and $b$) on has access to many Higgs decay
channels. This is exemplified in Fig.~\ref{allH} where the Higgs cross sections
at $\sqrt s=14$ GeV are shown together with the Higgs branching ratios; the 
(not yet excluded in summer) mass range  115 GeV$\lsim \! M_H \!\lsim \!130$ GeV
is highlighted. We summarise below some of the information available on the
determination of the Higgs properties, relying on the ATLAS and CMS technical
design reports  \cite{atlastdr,CMSTDR} and on some other analyses which  need
eventually to be updated.

\subsection{Mass, width and couplings of the SM Higgs}

The ease with which information can be obtained for the Higgs profile clearly
depends on the mass. The accuracy of the mass determination is driven by the
$H\to \gamma \gamma$ and  $H\to ZZ^* \to 4\ell$  modes for a light Higgs  and,
in fact, is expected to be accurate at one part in 1000. This is particularly
true  for the mass range 115 GeV $\lsim M_H \lsim 130$ GeV. 

\begin{figure}[!h]
\vspace*{-4mm}
\begin{center}
\includegraphics*[width=4.5cm,height=4.5cm]{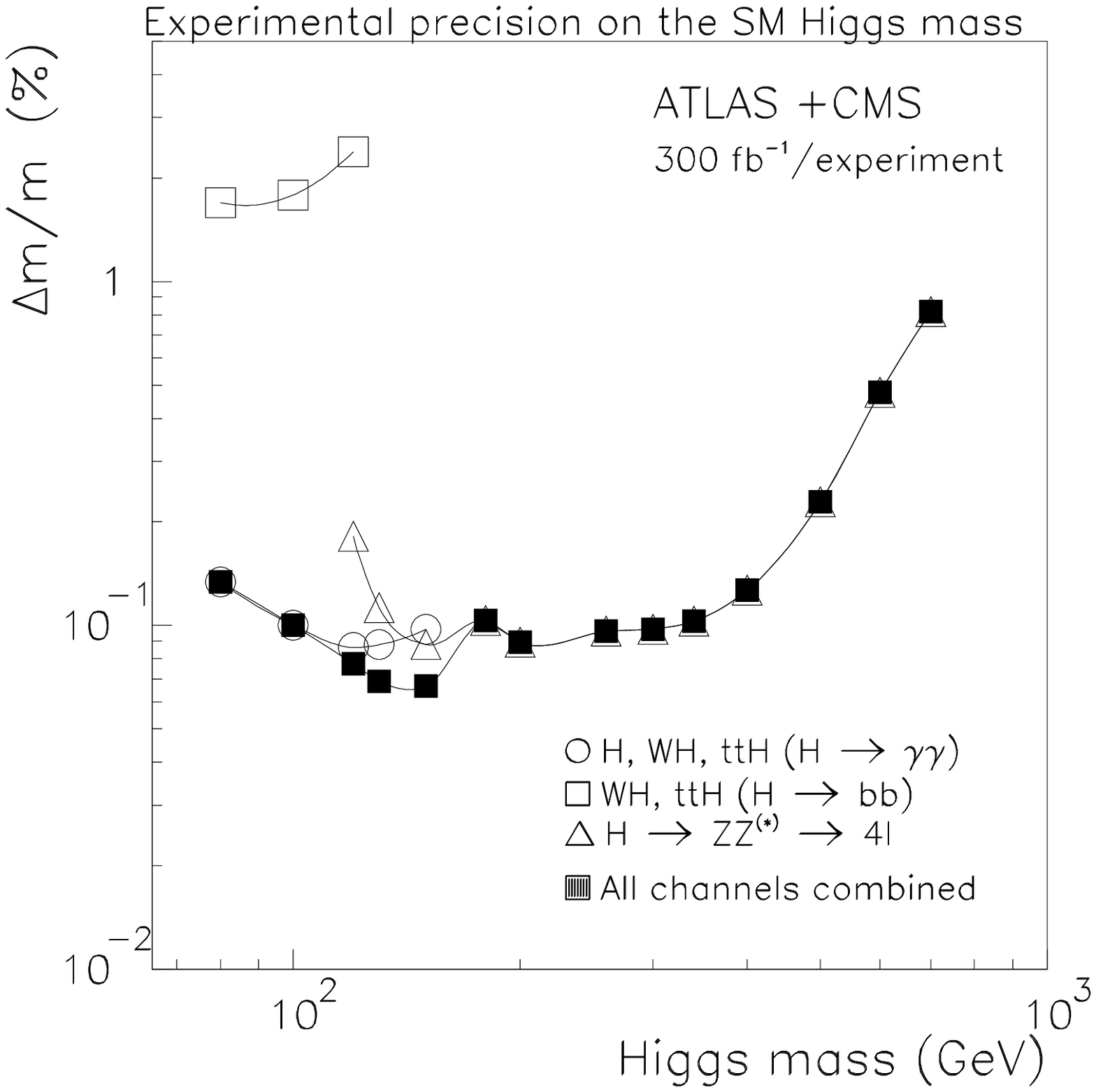}\hspace*{-1mm}
\includegraphics*[width=4.5cm,height=4.5cm]{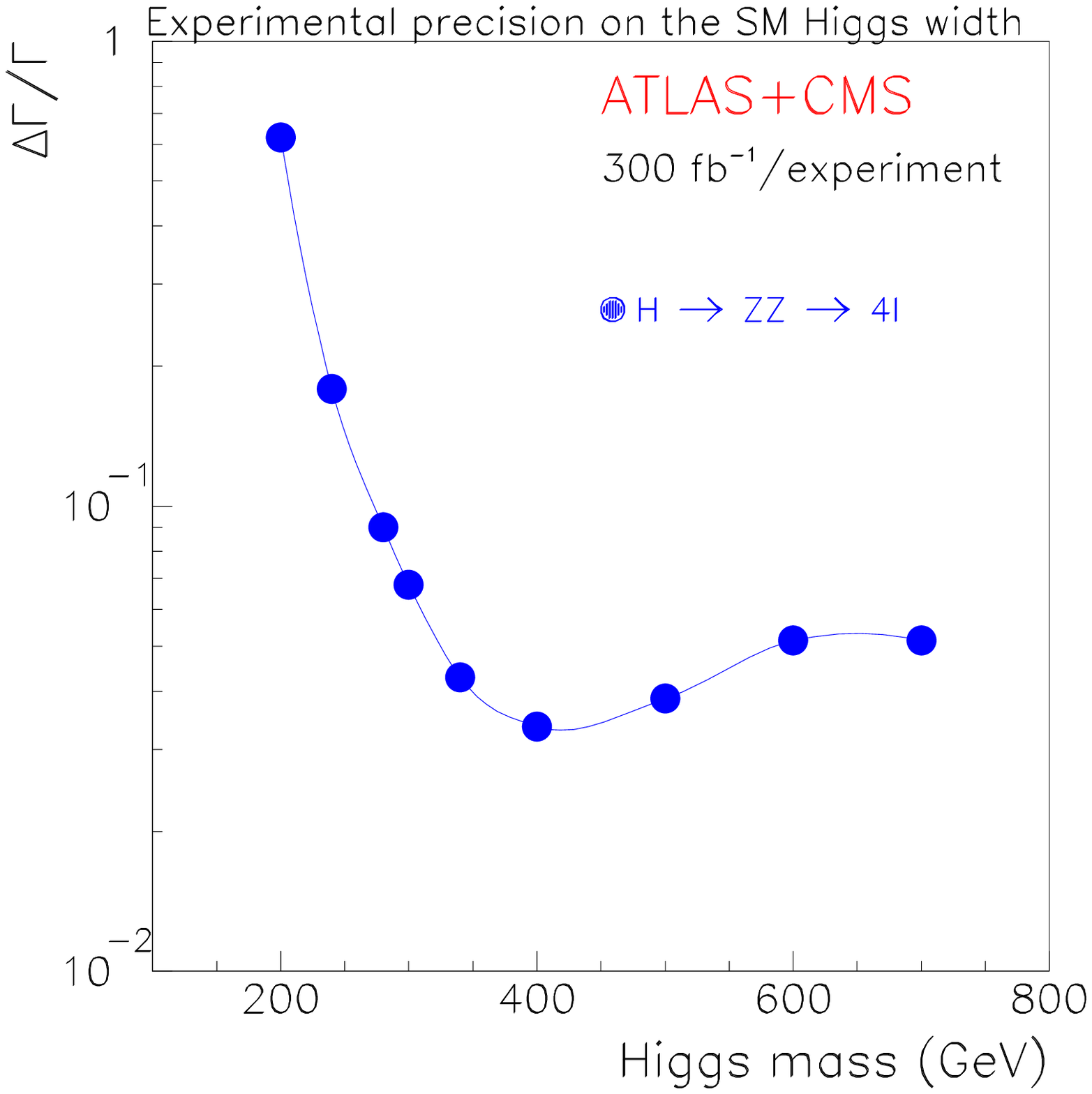}
\end{center}
\vspace*{-4mm}
\caption{Precision possible for the mass (left) and total width
(right) measurements for the SM Higgs for ${\cal L}=300$ fb$^{-1}$ 
combining ATLAS and CMS~\cite{atlastdr}.} 
\label{Hmass:preclhc}
\vspace*{-.1mm}
\end{figure}

Using the same process, $H \to ZZ \to 4\ell^\pm$, the Higgs total decay width
can be measured for $M_H\! \gsim\! 200$ GeV when it is large enough. For the 
mass range 115 GeV$\lsim M_H \lsim 130$ GeV, the Higgs total width is so small,
$\Gamma_H \lsim 5$ MeV,  that it cannot be  resolved experimentally. Only 
indirect means allow thus to measure the total Higgs decay width.

\begin{figure}[!h]
\vspace*{-.2mm}
\begin{center}
\includegraphics*[width=7cm,height=5cm]{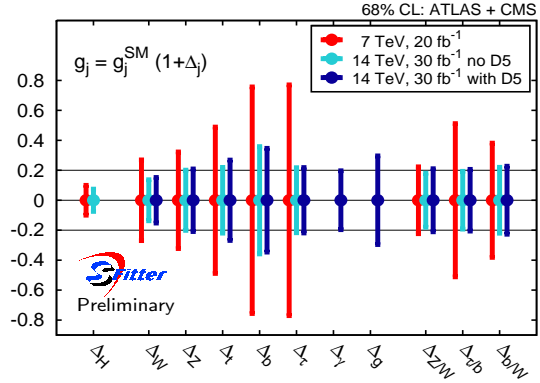}
\end{center}
\vspace*{-5mm}
\caption{Relative precision of the couplings of a 125 GeV Higgs boson at the 
LHC for some assumptions on the energy and luminosity; from
Ref.~\cite{Sfitter}.}
\label{Hcoup:preclhc}
\vspace*{-1mm}
\end{figure}

The determination of the Higgs couplings and the test of  their  proportionality
to the masses of fermions/gauge bosons, is absolutely essential for checking the
Higgs mechanism of EWSB. Ratios of Higgs couplings squared can be determined by
measuring ratios of production cross sections times decay branching fractions 
and accuracies at the 10--50\% can be obtained in some cases \cite{Dieter}.
However, it has been shown in Ref.~\cite{Dieter} that with some theoretical
assumptions, which are valid in general for multi-Higgs doublet models, the
extraction of absolute values of the couplings rather than just ratios of the
couplings, is possible by performing a fit to the observed rates of Higgs
production in different channels. For Higgs masses below 200~GeV  they find
accuracies of order $10$--$40\%$ for the Higgs couplings after several years of
LHC running.  A recent analysis of the determination of the Higgs couplings has
been performed in Ref.~\cite{Sfitter} for various scenarios of LHC energies and
luminosities.  The results are displayed in Fig.~\ref{Hcoup:preclhc} for a 125
GeV Higgs boson and as can been seem, accuracies of at most 20\% can be achieved. 

\begin{figure}[!h]
\vspace*{-1.3cm}
\begin{center}
\epsfig{figure=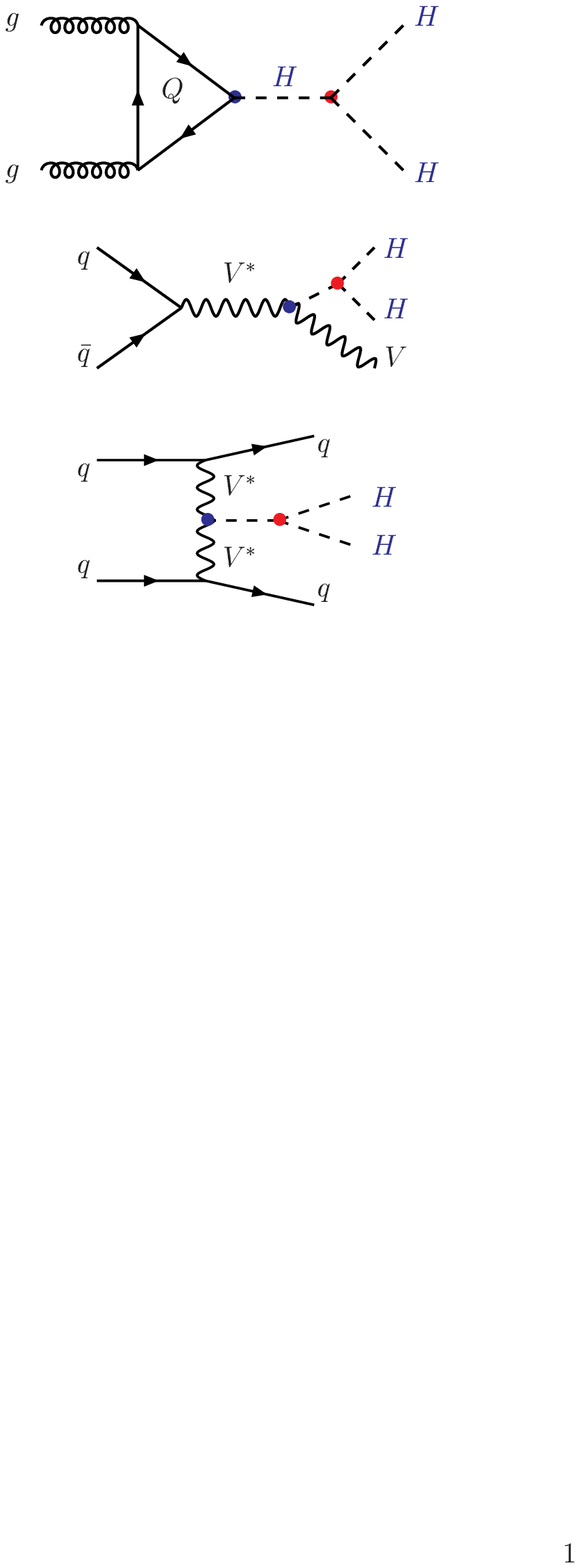,width=2.5cm}\hspace*{-1mm}
\epsfig{figure=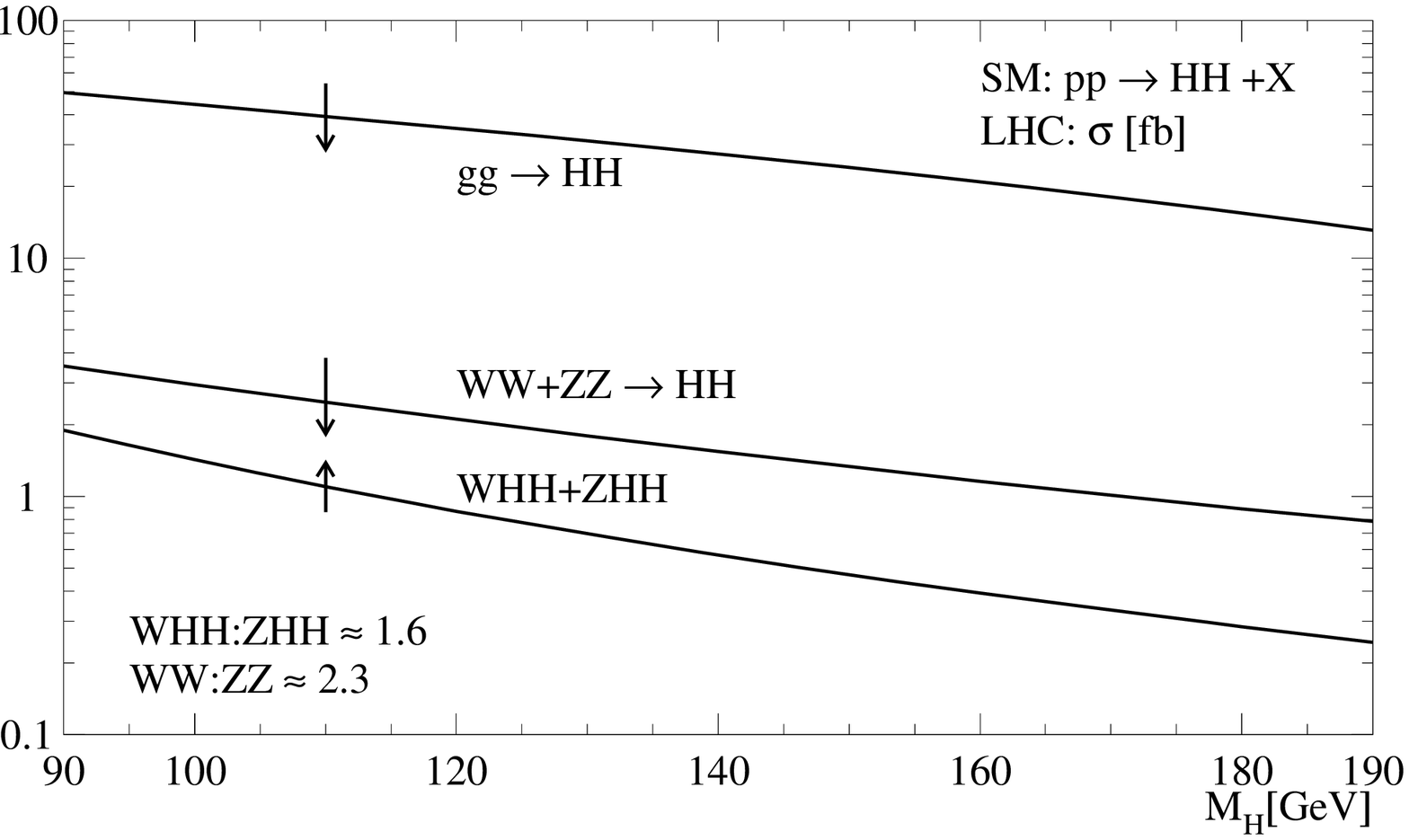,width=4.2cm,height=4.4cm}\hspace*{3mm}
\includegraphics[width=4.2cm,height=4.2cm]{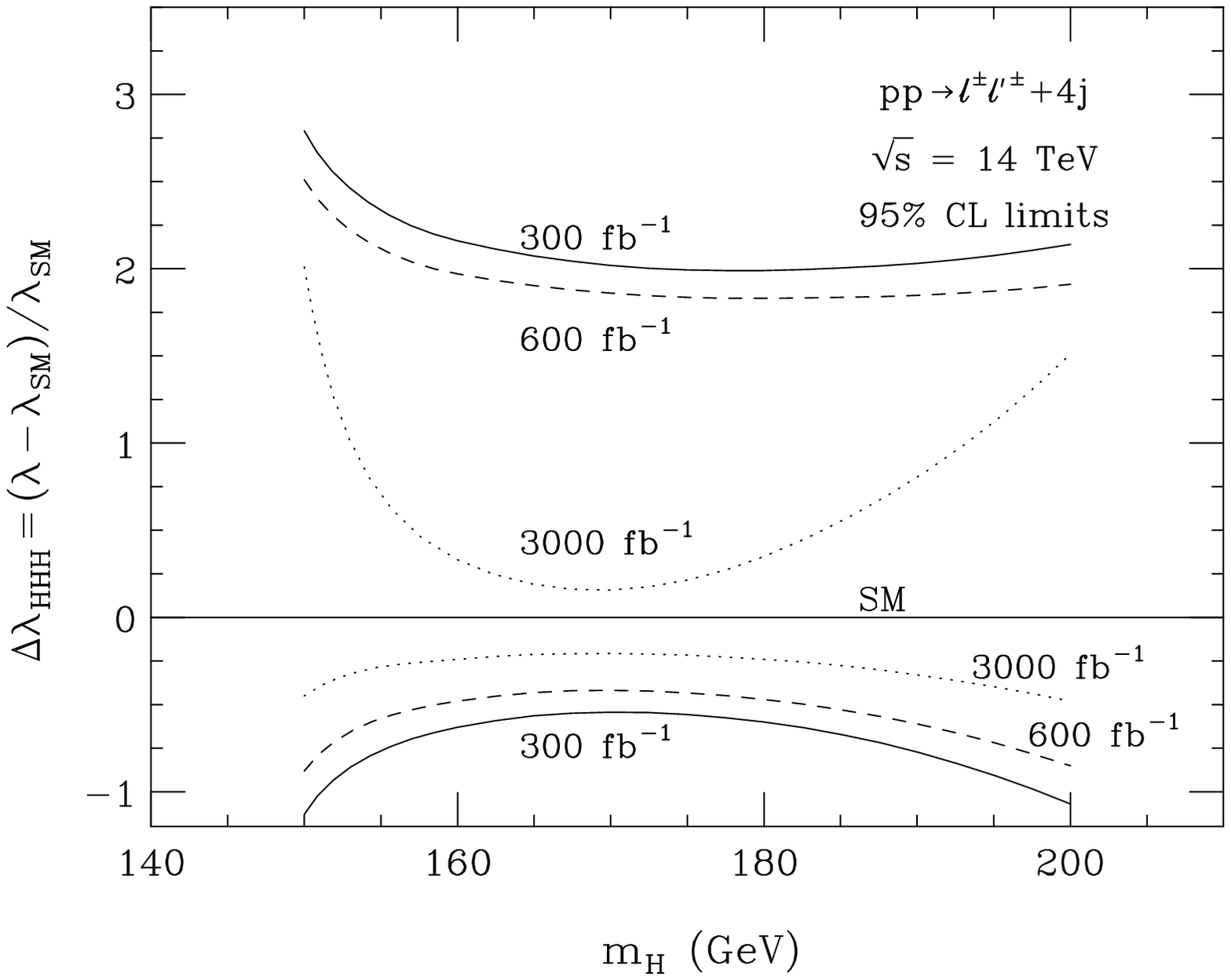} 
\end{center}
\vspace*{-5mm}
\caption{The main channels for double Higgs production at hadron colliders 
(left) and their cross sections at the LHC with $\sqrt s=14$ TeV (center) 
\cite{HH-LHC}; the statistical significance in the measurement of the triple 
Higgs coupling for various luminosities in the channel $pp \to \ell \ell '+4j$ 
\cite{HH-WW} (right).}
\label{Fig:HH}
\vspace*{-1mm}
\end{figure}

The trilinear Higgs boson self--coupling $\lambda_{HHH}$ is too difficult to be
measured at the LHC because of the smallness of the cross section in the main
double production  channel  $gg \! \to\!  HH$ and, a  fortiori, in the $qq\! 
\to\! qqHH$ and $q\bar q\! \to\!  HHV$ channels \cite{HH-LHC}; see
Fig.~\ref{Fig:HH}. In addition, the QCD backgrounds are formidable. A parton
level analysis has been  performed in the channel $gg\to HH \to (W^+W^-)(W^+W^-)
\to (jj \ell \nu) (jj \ell \nu)$ and $(jj \ell \nu) (\ell \ell \nu \nu)$ with
same sign dileptons, including all the relevant large backgrounds \cite{HH-WW}.

The statistical significance of the signal is very small, even with an extremely
high luminosity,  and one can at most set rough limits on the magnitude of the
Higgs self-coupling;  Fig.~\ref{Fig:HH}.  However, for a Higgs with a mass below
130 GeV, BR$(H\to WW^*)$ is too small and the channel $H\to b\bar b$ is affected
by a too large QCD background to allow for a measurement of the triple Higgs
coupling at the LHC.

Thus, for a very accurate and unambiguous determination of the Higgs couplings,
clearly an $e^+e^-$ linear collider \cite{DCR} would be more
suited.\vspace*{-3mm}

\subsection{Determination of the Higgs spin-parity}

One would  also need to determine the spin  of the Higgs boson and further
establish that the Higgs is a CP even  particle.  The observation of the decay
$H\!\to\! \gamma \gamma$ rules out the J=1 possibility.  Information on the CP
properties can be obtained by studying  various kinematical distributions such
as the invariant mass  of the decay products and  angular correlations among
them, as well distribution of the production processes. A large amount of work 
has been done on how to  establish, at different colliders, that the Higgs boson
is indeed   ${\rm J^{PC} = 0^{++}}$ state \cite{CPV-MSSM}.  Most of the 
analyses/suggestions for the LHC emanate by translating the strategies devised
in the case of an $e^+e^-$ collider and we will give only a few examples here. A
more detailed discussion can be found in Ref.~\cite{Djouadi:2005gi}.

\begin{figure}[!h]
\vspace*{-4mm}
\begin{center}
\mbox{\hspace*{-2mm}
\includegraphics*[width=4.3cm,height=4.8cm]{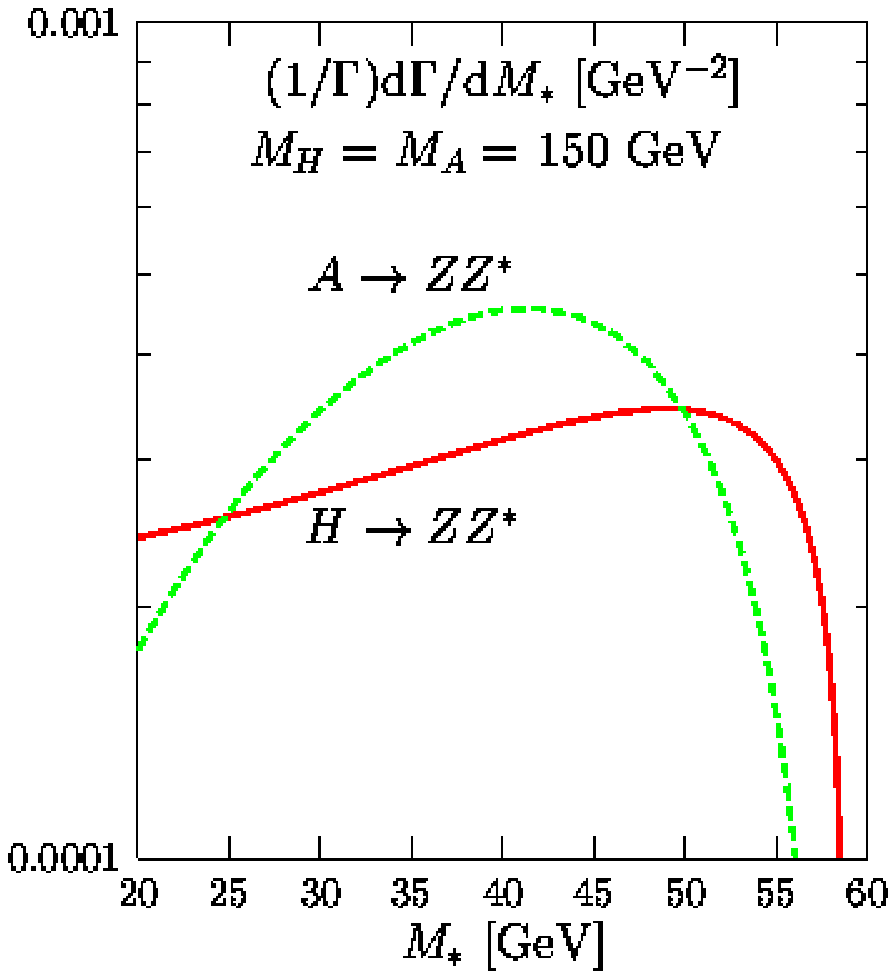}\hspace*{-2mm}
\includegraphics*[width=4.3cm,height=5cm]{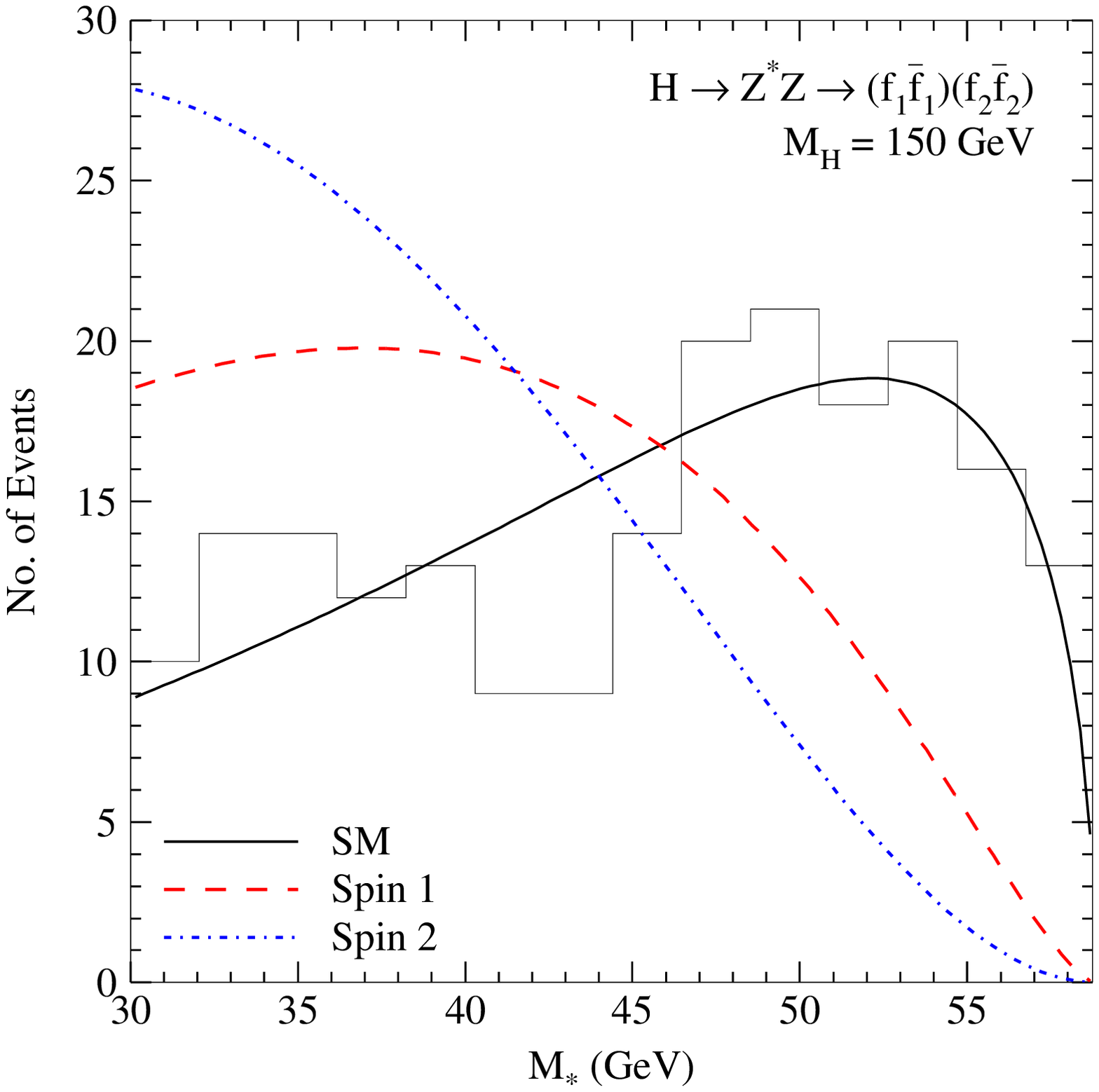}\hspace*{-3mm}
\includegraphics*[width=4.3cm,height=4.8cm]{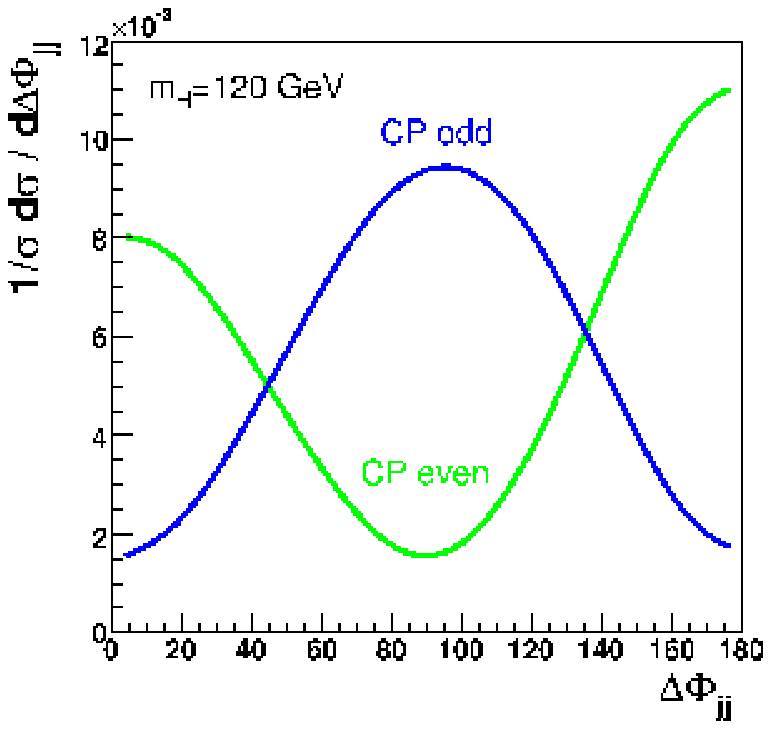}
}
\end{center}
\vspace*{-3mm}
\caption{Left: dependence on the CP nature of the Higgs boson of  the threshold
 behaviour of the distribution in $M_{Z^*}$ in $H\! \to\! ZZ^*\! \to \!4\ell $ 
decay~\cite{Bargeretal}. Center: Higgs spin determination via the threshold
behaviour of the distribution in $M_{Z^*}$  for the $H \! \to\!  ZZ^*\! \to \!
\ell$ decay~\cite{Choietal}. Right: Azimuthal angle  distribution for the two 
jets produced in association with the Higgs for CP-even and CP-odd cases
\cite{Hankele}.}
\label{Fig:CP}
\vspace*{-.1mm}
\end{figure}

One example is to study the threshold behaviour of the $M_{Z^*}$ spectrum in the
$H\! \to\! Z Z^{(*)}\! \to \to 4\ell$ decay for $M_H \lsim 2M_Z$
\cite{Bargeretal,Choietal}. Since the relative fraction of the longitudinally to
transversely polarised $Z$ bosons varies with $M_{Z^*}$, this distribution is sensitive
to both the spin and the CP nature of the Higgs. This is seen in
Fig.~\ref{Fig:CP}  where the behaviors for a CP-even and CP-odd states (left) 
and for different spins (center) are shown.

Another very useful diagnostic of the CP nature of the Higgs boson is the
azimuthal distribution between the decay planes of the two lepton pairs arising
from the $Z, Z^{(*)}$ bosons coming from the Higgs  decay
\cite{CPV-MSSM,Bargeretal,Choietal,CP-others}. Alternatively, one can study the
distribution in the azimuthal angle between the two jets produced in 
association with the Higgs produced in  the vector boson fusion process or in
gluon fusion in Higgs plus jet events~\cite{CP-VBF,Hankele}. Figure~\ref{Fig:CP}
(right) shows the azimuthal angle distribution for the two jets produced in
association with the Higgs, for the CP--even and CP--odd cases. With a high
luminosity of $300$ fb$^{-1}$, it should be possible to use these processes
quite effectively.

However, one should recall that any determination of the Higgs CP properties
using a process  which involves the couplings to massive gauge bosons is
ambiguous as only the CP even part of the  coupling is projected out. The
couplings of the Higgs boson with heavy fermions offer therefore the best
option. $t\bar t$ final states produced in the decay of an inclusively produced
Higgs or in associated production with a Higgs boson   can be used to  obtain
information on the CP nature of the $t\bar t H$ coupling, mainly  through
spin-spin correlations~\cite{CP-tt}. However, in the latter case, one  has
to consider the decay $H\to b\bar b$ which seems to be a difficult channel as
discussed earlier.  Another  approach which has been advocated ~ is to use
double-diffractive  processes with large rapidity gaps where only scalar Higgs
production is  selected \cite{CP-diff}. 

Most of the suggested measurements should be able to verify  the CP nature of a
Higgs boson  when the full luminosity of $300\,$fb$^{-1}$ is collected at the
LHC or even before, provided the Higgs boson is a CP eigenstate. However, a
measurement of the CP mixing is much more difficult, and a combination of
several different observables will be essential. In particular,  the subject of
probing CP mixing reduces more generally to the  probing of the anomalous $VV H$
and $t \bar t H$ couplings, the only two cases where such study can even be
attempted at the LHC and this becomes a precision measurement which is best
performed at $e^+e^-$ colliders as discussed for instance in
Refs.~\cite{DCR,anomalous}.\vspace*{-3mm}

\subsection{Measurements in the MSSM}

In the decoupling regime when $M_A \gg M_Z$ (or the antidecoupling regime for
small $M_A$), the measurements which can be performed  for the SM Higgs boson
with a mass $\!\lsim\! 115$--135 GeV will also be possible for the  $h(H)$ 
boson.  Under some assumptions and with 300 fb$^{-1}$ data,  coupling
measurements would allow to distinguish an MSSM from a SM Higgs particle at the
$3\sigma$ level for $A$ masses up to $M_A=$300--400 GeV \cite{Dieter}.  

The heavier Higgs particles $H,A$ and $H^\pm$ are accessible mainly in the $gg\!
\to\! b \bar b\!+\! H/A$ and $gb\!  \to\! H^\pm t$ production channels at large
$\tb$, with the decays $H/A\! \to\! \tau^+ \tau^-$ and $H^+\! \to\!  \tau^+
\nu$. The Higgs masses cannot be determined with a very good accuracy as a
result of the poor resolution. However, for $M_A \lsim 300$ GeV and with high
luminosities, the $H/A$ masses can be measured with a reasonable accuracy by
considering the rare decays $H/A \to \mu^+ \mu^-$  \cite{CMSTDR}. The
discrimination between $H$ and $A$ is though  difficult as the masses are close
in general and the total decay widths large \cite{intense}. 

\begin{figure}[h]
\vspace*{-6mm}
\begin{center}
\vskip 0.1 in
\includegraphics[width=70mm,height=45mm]{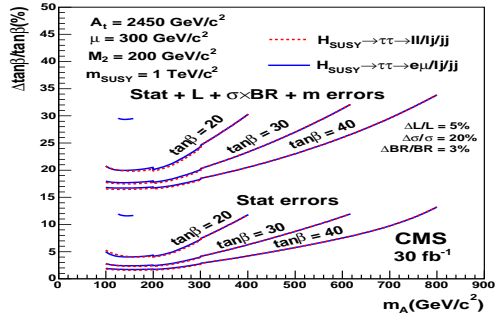}
\end{center}
\vspace*{-4mm}
\caption{The uncertainty in the measurement of $\tb$ in  $gg\!\to\! H/A\!+\!b
\bar b$ with the combined $H/A \!\to\! \tau \tau$ decays at CMS with 30 
fb$^{-1}$ data;  from Ref.~\cite{Sasha-tb}
The three lower curves show the uncertainty when  only statistical errors are
taken into account, while the upper curves include a 30\% theory+experimental
uncertainty.}
\vspace*{-3mm}
\label{Sasha}
\end{figure}

There is, however, one very important measurement which can be performed in
these channels. As the production cross sections above are all proportional to
$\tan^2\beta$ and, since the ratios of the most important decays fractions are
practically independent of $\tb$ for large enough values, one has an almost
direct access to this parameter.   A detailed simulation shows that an  accuracy
of $\Delta \tb/\tb \sim 30\%$  for $M_A\!\sim\! 400$ GeV and $\tb\!=\!20$ can be
achieved with 30 fb$^{-1}$ data \cite{Sasha-tb} as exemplified in
Fig.~\ref{Sasha}. 

\section{A glimpse into the future}

We are approaching the moment of truth where either a Higgs particle is
observed, thereby confirming the widely assumed scenario of spontaneous
electroweak  symmetry breaking by  a scalar field that develops a vacuum
expectation value,  or a revolution in particle physics will take place.  In
fact, a few months after this conference, the ATLAS and CMS collaborations have
collected  a sufficient amount of data to be sensitive to a SM--like Higgs
particle in the most interesting mass range, $M_H\lsim 130$ GeV and, in December
2001, the two  collaborations released preliminary results of their Higgs
searches  on almost  5~fb$^{-1}$ data per experiment. They reported an excess of
events over the SM background  at a mass of $\sim$ 125~GeV \cite{evidence},
which  offers a tantalising indication  that  a first sign of the Higgs boson
might be emerging. A Higgs particle with a mass of $\approx 125$ GeV would be 
a  triumph for the SM as the high--precision electroweak data are hinting since
many years to a light Higgs boson, $M_H \lsim 162$ GeV at 95\%CL. The ATLAS and
CMS results, if confirmed, would also have far reaching consequences for
extensions of the SM and, in particular, for supersymmetric theories
\cite{implications}. These results would close the search chapter of the Higgs
saga which lasted more than two decades and will open a new one: the precise
determination of the Higgs boson profile and the unraveling of the EWSB
mechanism. We hope that this new chapter will not last as long as the search 
chapter. \bigskip

\noindent {\bf Acknowledgements:} We thank the organizers of Lepton Photon 2011
in Mumbai, in particular Rohini Godbole and Naba Mondal,  for the perfect
organization of this highly special event in such a highly special place
as well as for their patience.

\end{document}